\documentclass[12pt,a4paper]{article}

\usepackage{epsfig}
\usepackage{amstex}

\def\nostrocostrutto#1\over#2{\mathrel{\mathop{\kern 0pt \rlap 
  {\raise.2ex\hbox{$#1$}}}
  \lower.9ex\hbox{\kern-.190em $#2$}}}
\def\lsim{\nostrocostrutto < \over \sim}   

\newcommand{\eref}[1]{(\ref{#1})}      
\newcommand{\pp}{{$\pm$}}
\newcommand{\N}{\overline{\mathcal N}}
\newcommand{\labl}[1]{\label{#1}}

\catcode`@=11
\newcount\@tempcntc
\def\@citex[#1]#2{\if@filesw\immediate\write\@auxout{\string\citation{#2}}\fi
  \@tempcnta\z@\@tempcntb\m@ne\def\@citea{}\@cite{\@for\@citeb:=#2\do
    {\@ifundefined
       {b@\@citeb}{\@citeo\@tempcntb\m@ne\@citea\def\@citea{,}{\bf ?}\@warning
       {Citation `\@citeb' on page \thepage \space undefined}}%
    {\setbox\z@\hbox{\global\@tempcntc0\csname b@\@citeb\endcsname\relax}%
     \ifnum\@tempcntc=\z@ \@citeo\@tempcntb\m@ne
       \@citea\def\@citea{,}\hbox{\csname b@\@citeb\endcsname}%
     \else
      \advance\@tempcntb\@ne
      \ifnum\@tempcntb=\@tempcntc
      \else\advance\@tempcntb\m@ne\@citeo
      \@tempcnta\@tempcntc\@tempcntb\@tempcntc\fi\fi}}\@citeo}{#1}}
\def\@citeo{\ifnum\@tempcnta>\@tempcntb\else\@citea\def\@citea{,}%
  \ifnum\@tempcnta=\@tempcntb\the\@tempcnta\else
   {\advance\@tempcnta\@ne\ifnum\@tempcnta=\@tempcntb \else \def\@citea{--}\fi
    \advance\@tempcnta\m@ne\the\@tempcnta\@citea\the\@tempcntb}\fi\fi}
\catcode`@=12

\begin{document}

\setcounter{page}{0}
\thispagestyle{empty}

\title{Low and High Energy Limits \\ of Particle Spectra in QCD Jets} 

\author{Sergio Lupia\thanks{E-mail address: lupia@mppmu.mpg.de}  
\ and \  Wolfgang Ochs\thanks{E-mail address: wwo@mppmu.mpg.de}}

\date{{\normalsize\it Max-Planck-Institut 
f\"ur Physik, (Werner-Heisenberg-Institut) \\
F\"ohringer Ring 6, 80805 M\"unchen, Germany}}

\maketitle
\thispagestyle{empty}

\begin{abstract}
Charged particle energy spectra in $e^+e^-$ annihilation are compared with the
analytical predictions from the QCD  evolution equation in the 
Modified Leading Log Approximation. 
With the nonperturbative initial condition shifted down to threshold as 
suggested by the  Local Parton Hadron Duality picture a good description of
the data from the lowest   
 up to highest available energies results. The 
two essential parameters in this approach 
are determined from a moment analysis. 
The sensitivity of the fit to the running of $\alpha_s$  and to 
the number of active flavours (including a light gluino) is demonstrated. 
For very high energies the theory predicts a scaling behaviour in 
certain rescaled variables
(``$\zeta$-scaling''). The data show an approximate behaviour of this type 
in the present energy range and come close to the predicted asymptotic 
scaling function for the small particle energies.
\end{abstract}

\vspace{-15.5cm}

\rightline{MPI-PhT/97-26}
\rightline{April 14th, 1997}

\newpage

\section{Introduction}
The  perturbative QCD describes well the properties of hard processes. 
The problem of the soft limit of the perturbation theory and the transition
to the hadronic final state, however, is still not solved at a fundamental
level. The  studies of the global properties of the hadronic final states 
gave support to 
the idea that the colour confinement mechanism should be rather soft  
 \cite{BAS,DKMTbook}. Meanwhile, the close similarity between hadronic and
partonic final states has been found for a large variety of observables,
provided the parton cascade in the perturbative calculation is evolved down
to rather low virtualities of the order of a few hundred MeV; this
phenomenon is called ``Local Parton Hadron Duality" (LPHD) \cite{LPHD} 
(for a recent review, see \cite{ko}).

An interesting prediction concerns the 
momentum spectrum of the particles in a jet at very high energies, namely 
the approximately Gaussian distribution
in the variable $\xi=\ln (P_{\mathrm{jet}}/p_{\mathrm{hadron}})$, 
the so-called ``hump backed plateau" \cite{DFK,MUEL,FW}, 
with a suppression of low momentum
particles following from the soft gluon interference \cite{ef,ahm1}.

Within the LPHD picture, not only the very high energy behaviour of the
spectrum is considered. Rather, one starts from the initial condition near
threshold for the parton cascade where only one parton is present and
derives the distribution at higher energies using the appropriate QCD
evolution equation.
In this case
 -- apart from the overall normalization -- the theory has only two essential
parameters, the QCD scale $\Lambda$ in the running coupling (at the one loop
level) and a cut-off $Q_0$ in the transverse momentum of the gluon emission.
In the 
Double Logarithmic Approximation (DLA) one takes into account the leading
contributions from the collinear and soft singularities which dominate at
very high energies and determine the asymptotic behaviour of the observables. 
At present
energies the next-to-leading corrections of relative order $\sqrt{\alpha_s}$
are  important and are fully taken into account in the Modified Leading
Logarithmic Approximation (MLLA).
An explicit analytical expression for the particle energy distribution 
is found in this approximation for the
special case $Q_0=\Lambda$, the so-called limiting spectrum \cite{LPHD}.
Remarkably, the experimental momentum spectra 
in the PETRA  \cite{tasso} and 
LEP \cite{opal} energy range  gave support to the perturbative predictions
\cite{LPHD}, and this success has been continued recently to the 
higher energies at LEP-1.5 (\cite{aleph15,delphi15,opal15} and \cite{klo1}) and 
the TEVATRON~\cite{Korytov}. 

The energy evolution of the spectrum and its connection to the initial
condition at threshold is most conveniently studied using the moment
representation of the spectrum. 
Recently, the first such moment analysis, based on the analytical results
\cite{DKTInt},  has been
performed over the full $cms$ energy range available in $e^+e^-$
annihilation \cite{lo}. There are a number of advantages of the moment
analysis over the analysis of the spectrum itself: (a) there are analytical
formulae which keep the two essential parameters of the theory 
$Q_0$ and $\Lambda$ independent~\cite{DKTInt}, whereas explicit formulae for
the spectrum are only available for the limiting case $Q_0$ = $\Lambda$; 
(b) the moments evolve with
energy independently of each other. Their absolute size is determined by the
initial condition at the threshold of the process; 
(c) the moments of order $q\geq 1$ are independent of the
overall normalization and depend only on the two essential parameters
of the LPHD approach; (d) the difference between the theory with running
coupling and an artificial theory with fixed coupling can be studied
directly and (e) the flavour thresholds can be included in the theoretical
calculations. As the calculations of the moments include the soft part of the
spectrum, the mass effects have to be taken into account. 

Having studied the behaviour of the spectrum towards low energies it is also
 interesting to study the limiting behaviour of the spectrum for very
high energies as the theories typically predict  simple asymptotic
expressions or a characteristic scaling behaviour. Then one can investigate  
to what extent this behaviour appears  already  at available energies.
The asymptotic behaviour of the QCD jet is obtained in the DLA. 
There is indeed a
finite scaling limit in certain rescaled variables 
and the scaling function can be calculated \cite{dfk}.

The purpose of this paper is twofold. First, we present more details of our
previous moment analysis \cite{lo} (see also \cite{conf}) 
as well as further results concerning all points (a) -- (e) above. 
Secondly, we give the description of the shape of
the spectrum and discuss the approach to the asymptotic  limit
in the appropriate scaling variables.

\section{Theoretical description of particle spectra}

\subsection{Evolution equations}

The multiparticle properties of a QCD jet is 
conveniently described by the generating functional
$Z(\kappa, Q_0,\{u(k)\})$. The arguments are the hardness scale $\kappa$
of the jet, which is defined in terms of jet momentum $P$ and 
opening angle $\Theta$ by
\begin{equation}
\kappa=2 P \sin \frac{\Theta}{2} \approx P\Theta
\label{kappa}
\end{equation}
where the approximation holds for small angles; $Q_0$ is the lower
cut-off in the transverse momentum of the emitted parton
\begin{equation}
k_\perp \geq Q_0.
\label{cutoff}
\end{equation}
The inclusive densities can be
obtained by functional differentiation of $Z$ over the probing functions
$u(k)$ refering to parton momentum $k$. 
Our analysis is based on the evolution equation for the gluon jet as
discussed in \cite{DKTInt}. We rederive here the basic equations 
for the energy spectrum whereby we start from
the evolution equation of the generating functional and show the
various steps of the approximations. 

The evolution of the generating functional with the hardness scale
$\kappa$ is given by the
``Master Equation'' \cite{DKMTbook,do} 
\begin{multline}
\frac{d}{d  \ln \, \kappa} \: Z_A (\kappa,Q_0)  = 
\frac{1}{2}
\; \sum_{B,C} \; \int_0^1 \; dz \\
 \; \times \frac{\alpha_s (k_\perp^2)}{2 \pi} \: \Phi_A^B (z)
\left [Z_B (z\kappa, Q_0) \: Z_C ((1 - z)\kappa,Q_0) \: - \: Z_A
(\kappa, Q_0) \right ]. \labl{master}
\end{multline}
This equation
describes the evolution of the jet by the splitting  of the
primary parton $A$ into  partons $B$ and $C$ at
 reduced scales $z\kappa$ and $(1-z)\kappa$; 
$\Phi_A^B(z)$ denote the
DGLAP splitting functions. 
The integral over the secondary energy fraction $z$ 
has to respect the limits (\ref{cutoff}) for the
transverse momentum $k_\perp = z (1-z)\kappa $.

The Master Equation (\ref{master}) 
yields results complete within the MLLA at high energies. Furthermore,  the
energy conservation constraints  are taken into account.
Because of (\ref{cutoff}) the evolution starts at $\kappa  =  Q_0$.
Then the initial condition for solving this system of equations reads
\begin{equation}
Z_A (\kappa, Q_0; \{ u \})|_{\kappa \: = \: Q_0} \; = \; u_A
(k =  P)
\label{zinit}
\end{equation}
which means that there is only one parton in the cascade at threshold.

The solution of the evolution equation~\eref{master} together with the boundary 
condition~\eref{zinit} yields perturbative expansions in $\alpha_s$ in all 
orders, which for many observables can be resummed and exponentiate at high
energies.
The MLLA takes into full account  the leading and next-to-leading
contributions in the expansion of $\sqrt{\alpha_s}$ in the exponent 
at high energies.

To this accuracy one can actually neglect the energy recoil 
and replace the $1-z$ by 1 
in the arguments of $Z$ and furthermore
neglect the $z$-dependence of $Z$ and of $\alpha_s(k_\perp(z))$
in the regular $z$-integrals. With these simplifications one
obtains a simplified form of (\ref{master}), 
the MLLA master equation \cite{do}. 
  
The evolution equation for the single inclusive density in 
the secondary parton energy $k$ 
is obtained from $Z$ by functional differentiation
\begin{equation}
D(k,\kappa,Q_0)=\frac{\delta}{\delta u(k)} Z(\kappa, Q_0,\{u\})|_{u=1}.
\label{Dsing}
\end{equation}
After differentiation
(\ref{Dsing}) of the MLLA master equation \cite{do} 
we obtain the evolution of the parton densities in quark and
gluon jets 
\begin{eqnarray}
\frac{d}{d\ln\kappa} D_g(k,\kappa) &=& \int_x^1 \frac{dz}{z}
     \gamma_0^2(z\kappa) D_g(k,\kappa) \nonumber\\
       & \hbox{ }& \qquad +\gamma_0^2(\kappa) \left(
    -\frac{11}{12} D_g(k,\kappa) + \frac{T_R \, n_f}{3 N_C} 
    \left[2D_q(k,\kappa)-D_g(k,\kappa)\right] \right) \nonumber\\
\frac{d}{d\ln\kappa} D_q(k,\kappa)
                     &=& \frac{C_F}{N_C} \left(\int_x^1 
     \frac{dz}{z} \gamma_0^2(z\kappa) D_g(k,\kappa)
    -\frac{3}{4} \gamma_0^2(\kappa)  D_g(k,\kappa)
      \right)
\label{Devolve}
\end{eqnarray}
Here $x=k/P$, 
 $n_f$ and $N_C$ denote the number of flavours and colours
respectively, $C_F=\frac{N_C^2-1}{2N_C}$ and $T_R=\frac{1}{2}$.
The anomalous multiplicity dimension $\gamma_0$
is given in one-loop approximation by
\begin{equation}
\gamma_0^2(k_\perp)=\frac{2N_C\alpha_s(k_\perp)}{\pi} 
 =\frac{4N_C}{b \ln (k_\perp/\Lambda)}, \quad
b=\frac{11}{3} N_C - \frac{2}{3}n_f. \labl{gamz}
\end{equation}
The initial conditions
for these evolution equations follow from (\ref{zinit})
\begin{equation}
D_A(k,\kappa)|_{\kappa=Q_0}=\delta(k-P_A) \quad , \quad A = q, g
\label{initial}
\end{equation}
It is convenient to write these equations in logarithmic variables, namely
\begin{gather}
Y=\ln \frac{\kappa}{Q_0}\approx \ln \frac{P\Theta}{Q_0}, \qquad
\xi=\ln\frac{1}{x} = \ln\frac{P}{k}, \qquad \lambda=\ln \frac{Q_0}{\Lambda},
 \nonumber \\
\xi'= \ln\frac{zP}{k}=\xi+\ln\,z, 
\qquad y'=\ln\frac{zP\Theta}{Q_0}=Y+\ln\, z.
\label{logvar}
\end{gather}
Then, after redefining $D(k,P,\Theta)\to D(\xi,Y)$ and
$D(k,zP,\Theta)\to D(\xi',y')$ we obtain after
integration of (\ref{Devolve}) with (\ref{initial})\footnote{We use here the
same notation for both densities $D(k) = dn/dk$ and $D(\xi) = dn/d\xi = k
dn/dk$.} 
\begin{align}
\begin{split}
 D_g(\xi,Y) =  \delta(\xi) &+ \int_0^\xi d \xi' \int_0^{Y-\xi} dy'
     \gamma_0^2(y'+\xi') D_g(\xi',y'+\xi') \\
         \qquad &+\int_\xi^Y dy  \gamma_0^2(y) \left(
    -\frac{11}{12} D_g(\xi,y) + \frac{T_R \, n_f}{3 N_C}
    \left[2D_q(\xi,y)-D_g(\xi,y)\right] \right) 
\nonumber 
\end{split} \\
\begin{split}
 D_q(\xi,Y) =  \delta(\xi) &+ \frac{C_F}{N_C} \biggl(\int_0^\xi d\xi' 
      \int_0^{Y-\xi} dy' \gamma_0^2(y'+\xi') D_g(\xi',y'+\xi') \\
       \qquad &-\frac{3}{4} \int_\xi^Y dy  \gamma_0^2(y) D_g(\xi,y) \biggr).
\end{split}\label{dqxiy}
\end{align}
The double integral terms in these equations originate from the
singular parts of the splitting functions $\Phi(z) \sim 1/z$ and
represent the leading double logarithmic terms of the 
DLA, the single integrals include the MLLA corrections
from the finite parts of the splitting functions 
within the required approximation. In the DLA
the spectra in quark and gluon jets are related by
\begin{equation}
 D_q(\xi,Y)-\delta(\xi) = \frac{C_F}{N_C} (D_g(\xi,Y)-\delta(\xi))
\label{dla}
\end{equation}
Replacing $D_q$ by its leading order contribution 
$\frac{C_F}{N_C} D_g$ in the nonleading term of 
eq.~(\ref{dqxiy}) for $D_g$, 
 we obtain the integral evolution
equation for $D_g$
\begin{equation}
\begin{split}
 D_g(\xi,Y) =  \delta(\xi) &+ \int_0^\xi d \xi' \int_0^{Y-\xi} dy'
     \gamma_0^2(y'+\xi') D_g(\xi',y'+\xi') \\
       \qquad\qquad  \qquad & \qquad\qquad  \qquad
       -\frac{a}{4N_C} \int_\xi^Y dy  \gamma_0^2(y) D_g(\xi,y)
\label{gln}
\end{split}
\end{equation}
with $a=\frac{11}{3} N_C + \frac{2n_f}{3N_C^2}$. The corresponding
differential equation reads
\begin{multline}
 \biggl(\frac{\partial}{\partial \xi} +  \frac{\partial}{\partial Y}\biggr)
\frac{\partial D_g(\xi,Y,\lambda)}{dY} -
\gamma_0^2(Y) D_g(\xi,Y,\lambda) \\ = 
- a \biggl(\frac{\partial}{\partial \xi} +  \frac{\partial}{\partial
Y}\biggr)
\biggl(\frac{\gamma_0^2(Y)}{4N_C} D_g(\xi,Y,\lambda)\biggr).
\label{dif}
\end{multline}
For
$a=0$ it corresponds to the DLA.
In order to solve this equation one can introduce the Laplace
transform $(D\equiv D_g)$
\begin{equation}
D(\xi,Y)= \int_{\tau-i\infty}^{\tau+i\infty} \frac{d\omega}{2\pi i}
     e^{\omega \xi} D_\omega(Y),
\label{laplace}
\end{equation}
where the integral runs parallel to the imaginary axis to the right
of all singularities of the integrand in the complex $\omega$-plane.
Then $D_\omega(Y)$ fulfils the ordinary linear differential equation
\begin{equation}
\begin{split}
\left ( \omega + \frac{d}{d Y} \right ) \; \frac{d}{d Y} \:
D_\omega (Y, \lambda) & - \gamma_0^2(Y) 
D_\omega (Y, \lambda)  \\
& =  - a \: \left ( \omega + \frac{d}{dY} \right ) \;
\frac{\gamma_0^2(Y)}{4N_C} 
   \: D_\omega (Y, \lambda) \; .
\end{split}
\label{domeq}
\end{equation}
This is the basic equation used in \cite{DKTInt} to derive the parton
distributions and their moments for gluon jets. For quark jets the 
high energy approximation $D_q=\frac{C_F}{N_C}D_g$ is taken. 
The differential equation~\eref{dif} has also been analyzed in \cite{lo}. 

One approach towards a solution of this equation is based on the anomalous
dimension ansatz
\begin{equation}
D_{\omega}(Y,\lambda) = D_{\omega}(Y_0,\lambda) \exp \left( \int_{Y_0}^Y dy
\gamma_{\omega}[\alpha_s(y+\lambda)] \right)
\label{anom}
\end{equation}
which yields a differential equation for                                             
$\gamma_{\omega}$ with two solutions,
one of which dominating at high energies.
Alternatively, one can find the solution of (\ref{domeq})
directly in terms of hypergeometric functions \cite{DKTInt,DKMTbook}.

\subsection{Moments of parton distributions}

There are various advantages of studying the moments of the parton
distributions as discussed in the Introduction. Analytical predictions for QCD
jets have been presented in~\cite{FW,DKTInt}. 

\subsubsection{Definitions and evolution equation} 

The unnormalized moments ${\cal M}_q$ 
of the $\xi$-distribution are defined by 
\begin{equation}
{\cal M}_q(Y, \lambda) = \int_0^Y d\xi \xi^q D(\xi,Y,\lambda)
\end{equation}                              
with ${\cal M}_0$ equal to the average parton multiplicity $\N$. 
They are closely related to the Laplace transform 
 \begin{equation}
D_\omega (Y, \lambda)  = 
\int_0^Y \; d \xi e^{- \xi \omega} \: D (\xi, Y, \lambda), \label{dom}
 \end{equation} 
by 
 \begin{equation}
{\cal M}_q   =  (-1)^q \frac{\partial^q}{\partial\omega^q}
D_\omega(Y,\lambda)|_{\omega=0} \; .
 \label{xiq} 
 \end{equation} 
The moments $<\xi^q>$
of the $\xi$-distribution are then defined by 
\begin{equation}
<\xi^q(Y, \lambda)> = \frac{{\cal M}_q}{\N} \; . 
\label{momdef}
\end{equation} 

One also introduces the cumulant moments 
 $K_q(Y,\lambda)$; the moments of lowest order in $q$ are given by
\begin{alignat}{2}
K_1 &= <\xi> \equiv \bar \xi, \quad & K_2 &= \sigma^2 = <(\xi - \bar \xi)^2>, \nonumber \\
 K_3 &= <(\xi - \bar \xi)^3>, \quad & K_4 &= <(\xi - \bar \xi)^4> - 3 \sigma^4. 
\label{cumdefs}
\end{alignat}
The reduced cumulants are defined by
$k_q \equiv K_q/ \sigma^q$,
in
particular the skewness $s = k_3$ and the kurtosis $k = k_4$.
The cumulants $K_q$ of general order $q$ can be derived from the expansion:
\begin{equation}
\ln D_{\omega}(Y,\lambda) = 
\sum_{q=0}^{\infty} K_q(Y,\lambda) \frac{(- \omega)^q}{q!}
\label{cumulants}
\end{equation}
and therefore
\begin{equation}
K_q(Y,\lambda) =  \left( - \frac{\partial}{\partial \omega}
\right)^q \ln D_{\omega}(Y,\lambda) \biggl|_{\omega=0}.
\label{kqdef}
\end{equation}
At high energies one term of the type (\ref{anom}) dominates and one finds
\begin{equation}
K_q(Y,\lambda) = K_q(0,\lambda) +  
\int_{0}^Y dy \left( - \frac{\partial}{\partial \omega}
\right)^q \gamma_{\omega}[\alpha_s(y+\lambda,n_f)] \biggl|_{\omega=0}
\label{kq}
\end{equation}
where the evolution starts at $Y=0$ according to the LPHD picture.


\setcounter{equation}{25}

The moments can be obtained via eq.~\eref{xiq}  from the evolution equation 
for $D_\omega$ with the appropriate boundary conditions at threshold. 
These are obtained from the Master Equation~\eref{master} 
with \eref{zinit} and \eref{Dsing} as 
\begin{alignat*}{2}
\N(0)&=1, \quad &  \N'(0)&=0 \tag{25} \label{multinit} \\
<\xi^q(0)>&=0 \quad  & <{\xi^q}'(0)>& =0 \quad \text{for}\quad q\geq 1. 
\end{alignat*} 
On the other hand, the boundary conditions implied by 
 eqs.~\eref{dqxiy} and~\eref{gln} are different in case of the multiplicity
 $\N$ because of the 
approximations involved in this integral equation,
whereas they remain the same for the higher moments. 
For $\lambda>0$ we find 
\begin{equation} 
\N(0) = 1 \quad , \quad \N'(0) = - \frac{B}{\lambda} < 0 \; .
\label{boundmlla}
\end{equation} 
The corresponding result is also obtained for 
$D_\omega$ in \cite{DKTInt}. 
This yields a minimum of the multiplicity above threshold with an unphysical
value $\N(Y_m) < 1$. In  case of the limiting spectrum with $\lambda=0$ 
both the position in $Y$ and the value of the minimum of the multiplicity $\N$ 
tends to zero and one obtains the boundary conditions $\N(0) = 0$ and  
$\N'(0) > 0$\cite{DKTInt}. 
These problems can be traced back to the approximations in 
the second integrals
in ~\eref{dqxiy} and~\eref{gln}: the correct limits~\eref{multinit} 
would be obtained, for
example, using the modifications ($\epsilon$-terms) proposed in\cite{do}. 

It should also be noted that the limit $\lambda \to 0$ is only possible because
of these approximations. Otherwise, whereas the energy dependence of the
multiplicity  $\N \sim \exp (c \sqrt{Y+\lambda})$ has a smooth limit for $\lambda \to 0$, 
the absolute normalization (the prefactor) would diverge 
for $\lambda \to 0$. This can be seen, for example, in the DLA from
\eref{mllamult} with $B = 0$. It will be interesting to study the effect of
these approximations in more detail. 
The subsequent analysis is based on the approximate eqs.~\eref{gln} 
and~\eref{domeq} with the corresponding boundary conditions. 


\subsubsection{Moments for running coupling $\alpha_s$} 

The results obtained from the full solution of \eref{domeq} 
can be written for arbitrary parameters 
$Q_0$ and $\Lambda$  as:
\begin{equation}
< \xi^q > = \frac{1}{\N} \sum_{k=0}^q {d \choose k} ( N_1 L_k^{(q)}
+ N_2 R_k^{(q)} )
\label{moments:full}
\end{equation}
where $N_1$, $N_2$, $L_k^{(q)}$ and $R_k^{(q)}$ are known functions of
 $a$, $b$, $Y+\lambda$ and  $\lambda$ whose explicit expression depends on
 the order $q$ \cite{DKTInt} (see Appendix A1). For
the special case of the limiting spectrum, 
 where the two parameters  $Q_0$ and $\Lambda$ coincide (i.e. $\lambda = 0$),
the expressions simplify and all moments can be expressed in terms of
the
parameter $B \equiv a/b$
and the variable $z \equiv \sqrt{ 16 N_c Y / b }$. 
The general result for the  $q$-order moments is then the following \cite{DKTInt}:
\begin{equation}
\frac{<\xi^q>}{Y^q} = P_0^{(q)}(B+1,B+2,z) + \frac{2}{z}
\frac{I_{B+2}(z)}{I_{B+1}(z)} P_1^{(q)}(B+1,B+2,z)
\label{moments:LS}
\end{equation}
where $P_0^{(q)}$ and $P_1^{(q)}$ are polynomials
of order $2(q-1)$.
Expanding the Bessel functions and $P_i^{(q)}$, one obtains a series in
$1/\sqrt{Y}$. The leading and next-to-leading order results in this
expansion (see also \cite{FW}) determine the high energy behaviour; 
the remaining part of the series
however is still numerically sizeable at LEP energies (10\% contribution to
$\bar \xi$ and $\sigma^2$) and increases towards lower energies. We
therefore included the full result~\eref{moments:LS}, 
also for $P_i^{(q)}$ (for the explicit expressions, see Appendix A2).
The  average multiplicity of partons
is given in this approximation by:
\begin{equation}
\N_{LS} = \Gamma(B) \left( \frac{z}{2} \right)^{1-B} I_{B+1}(z).
\label{norm:LS}
\end{equation}

\subsubsection{Moments for fixed coupling $\alpha_s$} 

In this case the differential equations (\ref{dif}) and (\ref{domeq}) 
can be solved exactly \cite{lo}. For $D_\omega$  one finds
\begin{equation}
D(\omega,Y)= \left(\frac{\omega/2+\eta}{\tilde\omega}
  \sinh(\tilde\omega Y)
  +\cosh(\tilde\omega Y)\right)
   e^{-(\omega/2+\eta)Y}   \label{domfa}
\end{equation}
where $\tilde\omega =\sqrt{(\omega/2 - \eta)^2+\gamma_0^2}$.
Differentiation (\ref{xiq}) yields the moments in the form
\begin{equation}
\N_{fix} =  \left[ \cosh \left( \bar \gamma_0 Y \right) + 
\frac{\eta}{\bar \gamma_0}  \sinh \left( \bar \gamma_0  Y
\right) \right] \exp \left(  - \eta Y \right) 
\label{norm:fix:full}
\end{equation}
\begin{gather}
< \xi^q_{fix}>  =  \left[ A_q \cosh \left( \bar \gamma_0 Y \right)
+ B_q \sinh \left( \bar \gamma_0 Y \right) \right]
\frac{\exp \left(  - \eta Y \right) }{\N_{fix}}\label{xiqfix}\\
 \bar \gamma_0 \equiv \sqrt{\gamma_0^2 + \eta^2} \qquad
  \eta=\frac{a\gamma_0^2}{8N_C}=\frac{a\alpha_s}{4\pi}  \nonumber
\end{gather}
The coefficients $A_q,~B_q$  
are polynomials of order $q$ in $Y$,  given for $q \le 4$ 
in the Appendix A3. 
For  $\eta$ = 0, one obtains back the DLA results, especially 
$\N_{fix}=\cosh(\gamma_0 Y)$.

These results are obtained using the original boundary conditions, 
eq.~\eref{multinit}. As pointed out above, from the MLLA equation~\eref{gln} 
the boundary conditions~\eref{boundmlla} are obtained. 
We have also studied the results following from these other 
boundary conditions: in this case the moments would be
shifted towards larger $Y$ at most by 1 unit and the overall 
description of the data would become worse. 

At high energies the results~\eref{xiqfix} 
greatly simplify and the moments read: 
\begin{equation} 
< \xi^q_{fix} > \simeq \frac{A_q + B_q}{1 + \eta/\bar \gamma_0} 
\end{equation} 
i.e., they behave like 
$< \xi^q_{fix} > \sim Y^q$. For the cumulant moments the
leading terms cancel and 
\begin{equation}
K_{q,fix} \sim Y \; .
\label{kqlim}
\end{equation}
This can be easily seen by noting that at high energies 
eq.~\eref{domfa}  can be approximately written as 
\begin{equation} 
D(\omega,Y) \simeq \exp \left[ \gamma_{\omega} Y \right]  
\qquad 
\hbox{\rm with} \qquad 
\gamma_{\omega} =  - (\frac{\omega}{2} + \eta) + \tilde \omega
\end{equation}
Since the coupling in this case is frozen, the anomalous dimension
$\gamma_{\omega}$ does not
depend on $y$, the integral in  eq.~\eref{kq} becomes trivial and
eq.~\eref{kqlim} follows directly.

\section{Moment analysis}

\subsection{Determination of moments and mass effects}

In order to determine the moments one has to integrate the $\xi $-spectra 
over the full range. Here one faces
a problem for small momenta because of mass effects as already discussed in
\cite{lo}.
In the theoretical calculation the partons are treated as massless
with $k_\perp > Q_0$, therefore $\xi \equiv \xi_E = \ln 
\frac{P}{E} \le Y = \ln \frac{P}{Q_0}$, i.e. $\xi$ has an upper limit.
On the other hand, the experimental data usually 
refer to the distribution
in particle momentum $p$ or $\xi_p = \ln \frac{P}{p}$, and
$\xi_p$ is not limited from above.

For identified particles with known masses 
one can easily construct the energy distributions,
but there is no reliable prediction yet for the mass dependence 
of the identified particle spectra within the theoretical framework
considered here. 
Therefore we restrict ourselves to 
charged particles and we give them a common effective mass $m_h$. 
If this mass is taken equal 
to the cut-off $Q_0$, then for partons and charged particles there is the
same upper limit $\xi_E = Y$.
The hadron spectra in the jet from the primary parton $A$ are then calculated
from the parton distribution according to the LPHD hypothesis from
\cite{dfk1,lo}
\begin{equation}
E_h \frac{dn_A(\xi_E)}{dp_h} = K_h E_p \frac{dn_A(\xi_E)}{dp_p}
    \equiv K_h D_A^g(\xi_E,Y)
    \label{phrel}
\end{equation}
with $E_h = \sqrt{p_h^2+Q_0^2}$ and  $K_h$  an unknown normalization constant
to be fitted by the data.
Eq. \eref{phrel} leads to the correct relation for $p \sim E \gg Q_0$,
independent of the mass;
for small momenta it yields a finite value for the invariant density
$E \frac{dn}{d^3p}$ in agreement with the data 
\cite{lo} (see also the discussion on this point in \cite{klo2}).
As $Q_0$ has the meaning of a transverse momentum cut-off for partons, 
it could be thought of for hadrons 
as an effective transverse mass $m_T = \sqrt{m^2+p_\perp^2}$, 
 which is larger than the particle mass itself.

The effect  of introducing an effective particle  mass  
on the shape of the spectrum is shown in 
Fig.~\eref{tassoopal}. Here the inclusive charged particle 
$\xi$-spectra with
different mass assignments,
namely $Q_0=0$, $Q_0=m_\pi$ and $Q_0=270$ MeV, are shown. 
The latter value is suggested from the
moment analysis \cite{lo}, see below. 
The upper limits of $\xi$ corresponding to the above effective masses $Q_0$  
are also shown in the figures. 
As can be seen from these figures, 
the rescaling procedure is relevant in the soft region only, where 
the kinematical boundary becomes important. The effect is stronger at
the lower $cms$ energy of $\sqrt{s}$ =
14 GeV \cite{tasso} as compared to $\sqrt{s}$ = 91 GeV\cite{opal}. 
In the latter case the separation of curves in the measured range is hardly
visible. 

The moments $<\xi^q>$ are determined from the spectra $E dn/dp$ vs. 
$\xi_E$ after appropriate transformation of the measured $x_p$ 
spectra 
and therefore
depend on the chosen effective mass $m_h=Q_0$. 
For the unmeasured interval near $\xi_E \simeq Y$ (small momenta), we added an
extra-point with coordinates \{$(Y + \xi_{last})/2$, 
$D(\xi_{last})/2 \pm \sigma_D(\xi_{last}$)\}
to linearly interpolate between the last measured point of
coordinates \{$\xi_{last}$, $D(\xi_{last})\pm \sigma_D(\xi_{last}$)\} and the 
limit \{$Y$, 0\} imposed by kinematics.
The errors of  the moments are determined from the errors of $D(\xi_i)$ and 
the errors of the central values of $\xi_E$ in
each bin, taken as half the bin-size. 

Correspondingly, we 
obtain the multiplicity $\N_E$ as integral over $\xi_E$ of the 
full spectrum $E dn/dp$. As expected from Fig.~\eref{tassoopal},  
its difference to 
the usual particle multiplicity $\N_{ch}$ decreases with rising 
$cms$ energy, from 30\%  at $\sqrt{s}$ = 3 GeV to 10\% at LEP energy.
The MARK I data point at $\sqrt{s}$ = 4.03 GeV 
shows an anomalous decrease of up to 50\%, which may be partly 
related to charm thresholds effects. The result of this moment determination
using the effective mass $Q_0$=270 MeV is presented in  Table~1; this value of
$Q_0$ results from the fit discussed below. 

\subsection{QCD description of moments for running $\alpha_s$}

\subsubsection{Determination of the parameters $Q_0$ and $\Lambda$} 

The study of cumulant moments of the charged particle energy spectra 
allows for the first time the unconstrained 
determination of the two essential parameters which enter the theoretical
predictions, namely  $Q_0$ and $\Lambda$ (or $\lambda$) for the running
$\alpha_s$ model and $Q_0$ and $\gamma_0$ for the fixed $\alpha_s$ model. 
Let us start with the running $\alpha_s$ case with results 
shown in Fig.~\eref{test3r}, the analysis of the fixed $\alpha_s$
results, shown in Fig.~\eref{test3f}, runs parallel and will be discussed below. 

Fig.~\eref{test3r} shows the mean multiplicity, the
average value $\bar \xi_E$ and the dispersion 
$\sigma^2$ extracted from the experimental
data \cite{tasso,opal,aleph15,delphi15,opal15,slac,slac2,topaz,aleph,delphi,l3} 
as a function of the $cms$ energy  
for three different values of the parameter $Q_0$; 
the theoretical predictions for the cumulant moments using the given $Q_0$ 
still depend on $\Lambda$ (or $\lambda$). 
The  predictions are calculated for the number of flavours $n_f $ = 3. 
For the particle multiplicity we use 
\begin{equation} 
\N_E = c_1 \frac{4}{9} 2 \N_{part} + c_2  
\label{multtwopara}
\end{equation}
with arbitrary parameters $c_i$ and parton multiplicity $\N_{part}$ from 
eq.~\eref{norm:LS} (the factor $4/9$ is for the quark jet and 2 is for the two
hemispheres). The two normalization
parameters are determined to let the curve go through the lowest
and the highest energy data points. 
The parameter $c_1$ corresponds to the $K_h$ factor in eq.~\eref{phrel},
  whereas the additional parameter $c_2$ has been introduced to allow for a
  finite multiplicity at threshold as in~\eref{multinit}. 
 It is important to note that the higher moments ($q\geq 1$)
 describe the shape and do not depend on the
 normalization, so they are unaffected both by the systematic
experimental
 uncertainties of the overall normalization and by the theoretical 
uncertainties  associated with the $K_h$ factor. 
  
  The mean multiplicity data in Fig.~\eref{test3r},
 for each chosen $Q_0$, can be properly described by the theoretical
  predictions for any value of $\lambda$ in the range $0\leq \lambda \leq
4$. Let us stress that
  the theoretical predictions at parton level
   strongly depend on $\lambda$, but one can obtain in all cases 
good fits of the experimental data 
by adjusting the free parameters $c_i$ in
  eq.~\eref{multtwopara}. Therefore
 the mean multiplicity data
alone do not  determine the parameter $\lambda$.
Looking now at the first moment $\bar \xi$, we observe that 
for each chosen $Q_0$ a suitable value of $\lambda$ can be found 
which provides a good
description of the data. In order to fix both parameters, one has then to
include
the dispersion $\sigma^2$. 
As can be seen from the figure, a  lowering of $Q_0$ 
 shifts both the $\bar \xi_E$ and the $\sigma^2$ data downwards. 
 On the other hand,
an increase  of $\lambda$ yields lower values 
 for $\bar \xi_E$ but larger values for $\sigma^2$, as one can see 
 by inspection of the  first two terms in the expansion of
$\bar \xi_E$ and $\sigma^2$ in $\lambda$ \cite{lo}. 

The parameters  $Q_0$ and $\lambda$ are determined from a
 $\chi^2$-minimization.
 To compute the $\chi^2$ we have used the first 4 moments ($q\geq 1$) of
  the inclusive energy spectra for charged particles, but not the 
  mean multiplicity because of its larger  systematic
  errors and the need of two more normalization parameters for its
theoretical description. 
  The  minimum of $\chi^2$ 
 is obtained for the limiting spectrum
($\lambda \to$ 0), and the parameters are estimated as: 
\begin{equation}
Q_0 \simeq \Lambda \simeq 270 \ \hbox{MeV}. 
\label{qnull}
\end{equation}

The minimum value of the $\chi^2$/$d.o.f.$, neglecting 
  the correlations among the cumulants of different order,  
  is found to be 1.8 (with about 70 $d.o.f.$). So we do not obtain a ``perfect" fit
of the data, but considering the small errors of the lowest order moments
and the small number of parameters, a very satisfactory description 
of all moments over a large energy range is obtained. In view of the systematic
uncertainties of the fit we estimate the errors of the parameters from the
limiting case in which the theoretical curves miss all data points by 
about one standard deviation.  This yields the conservative estimate
 \begin{equation} 
 \Delta Q_0 \simeq \Delta \Lambda \simeq 20\ \  \hbox{MeV}. \label{error} 
 \end{equation}
Alternatively, this result can also be transformed into 
a limit on $\lambda=\ln(Q_0/\Lambda)$
 \begin{equation}
 \lambda\lsim 0.1. \label{lambdalim}
 \end{equation} 
Our result~\eref{qnull}  
is slightly larger than the $\sim$ 250 MeV obtained by the OPAL
Collaboration\cite{opal} from a fit of the measured $\xi$ distribution
to the limiting spectrum adjusting only one parameter. The small difference 
 results from the inclusion of lower energy data in our fit. 

\subsubsection{Discussion of the fits} 

The cumulant moments up to order $q=4$ together with the corresponding 
predictions for the limiting spectrum with $Q_0$ = 270 MeV and $n_f=3$ 
 are shown in Fig.~\eref{momentsflavour} by the solid lines. 
 The values of the moments and the 
corresponding theoretical predictions are given also in Table 1. 
A very satisfactory description is obtained in the full  $cms$ energy range
available;
small deviations in the first moment at very low energies are visible. This 
may signal some limitations of the
approximations involved, in particular, the simplified relation between
quark- and gluon-jets, in this region. 

It should be noted that the moments with $q\geq 1$ are determined by two
parameters only which actually almost coincide. 
This should be contrasted with the more conventional applications of
perturbative QCD to the particle or parton spectra. 
There one starts at a
finite energy $Y_0$  with a nonperturbative input distribution which in
general requires a set of unknown parameters and then evolves this
distribution to higher energies according to the predictions of perturbative
QCD. In terms of moments this would require
one adjustable value  $K_q(Y_0)$ for each moment at the initial energy
$Y_0$. 

In the application of LPHD one assumes the validity of the perturbative
formulae for the moments
down to small energy scales of order few hundred MeV, 
actually down to the threshold energy $Q_0$, 
where the distribution function is known to be simply the $\delta$-function 
~\eref{dqxiy}. In this limit all the higher moments 
are determined to be zero. 
Therefore the   compact description with only two parameters  
is a direct consequence of the assumption that the theoretical description
can be continued down to these low energies. A free adjustment of
the moments at higher energy $Y_0$ with a vertical shift of the curves in 
 Fig.~\eref{momentsflavour} would not improve the predictions essentially;
the limiting spectrum with the absolute normalization  at 
threshold gives indeed the best results. It should be noted that the
previous applications of the limiting spectrum to fit the $\xi$-spectra 
(for example \cite{LPHD,opal}) 
rely on the same assumption that the QCD evolution can be continued down
to the low scales $Q_0$ of few hundreds MeV where the initial condition is
introduced.

\subsubsection{Flavour dependence} 

The theoretical predictions shown so far were obtained with 3 active
flavours. 
A possible source of uncertainty in the theoretical formulae
is the number of active flavours to be used (see also \cite{lo,klo1}). 
In the predictions for cumulants \eref{kq}
 the number of flavours enters essentially through the running coupling
$\alpha_s(y,n_f)$. We neglect in the present discussion 
the additional explicit dependence at the percent level, 
which comes in through the
parameter $a$ at the next-to-leading order of the MLLA. 

The moments evolve at low energy with 3 active flavours 
and with 4 and 5 flavours after passing 
the respective thresholds.
The simplest approach would be to put the thresholds at the heavy quark 
masses, i.e., to increase $n_f$ by one at $\frac{\sqrt{s}}{2}=m_Q$
where $\sqrt{s} = 2 P$. 
However, 
let us recall that the argument of $\alpha_s$ is the transverse momentum 
$k_\perp$  and kinematics forces 
$k_\perp \le \frac{1}{4} \frac{\sqrt{s}}{2}$. This suggests moving
the thresholds to $\frac{\sqrt{s}}{2} = 4 m_Q$ 
(or towards even larger values, see for example \cite{Marciano}). 

In Fig.~\eref{momentsflavour} we show the predictions from the limiting spectrum 
with the inclusion of heavy flavours
 at the corresponding thresholds $\frac{\sqrt{s}}{2} = 4 m_Q$ 
(dashed lines). Above a heavy quark threshold the moments evolve 
according to \eref{kq} with the 
respective number of active flavours $n_f$ and match continuously to the
moments with $n_f-1$ active flavours below the threshold, in complete
analogy to the inverse coupling constant  $1/\alpha_s$\cite{altarelli}.
 We then write for the cumulants
 \begin{equation} 
 K_q(\frac{\sqrt{s}}{2}) = 
   K_q^{(n_f)}(\frac{\sqrt{s}}{2}) -  \sum_{i=4}^{n_f} \biggl( 
   K_q^{(i)}(4 m_f) - K_q^{(i-1)}(4 m_f) \biggr) \Theta (\frac{\sqrt{s}}{2}- 
   4m_f ) 
 \end{equation} 
Here $K_q^{(i)}(P_{jet})$ refers to the moment calculated with $i$ flavours
from threshold $Q_0$ up to jet momentum $P_{jet}$. 
 The comparison of the upper two curves in Fig.~\eref{momentsflavour} 
shows that the inclusion of the heavy
 quark thresholds does not modify dramatically the behaviour of the moments. 
 Therefore, a reasonable approximation of the 
experimental  data in the present energy range is obtained by taking into
account only 
three active flavours throughout the full energy range. 
This behaviour of the theoretical predictions can be easily  understood from 
 the representation of the moments~\eref{kq} 
 in terms of the anomalous dimension. Just above
threshold the contribution of the new flavour to the $y$-integral is negligible.
 Most primary gluons are still emitted at a smaller scale below the
new flavour threshold. 
On the other hand, an approximation with five flavours in the full energy range
would be a complete failure. 
 
 We have neglected here the differences in
 the light and heavy quark fragmentation. Whereas such effects occur in the
 fragmentation region, they are expected to be small in our application, as the
 soft gluon radiation is universal. 

\subsubsection{The effect of a light gluino} 

By exploiting 
 the flavour dependence one can also extract new information on the
 possible presence of additional light particles. 
 There has been considerable interest in the last years in the question whether
 there is a supersymmetric gluino with a small mass. In a recent
 summary~\cite{gluino} the gluinos in the mass range
 1$\frac{1}{2}$--3$\frac{1}{2}$ GeV are considered as absolutely excluded,
 whereas lighter gluinos are allowed, except for certain ranges of lifetime. 
In a recent study~\cite{alephgluino} of jet rates and jet 
 angular distributions in the reaction $e^+e^- \to 4$ jets, such a possibility
 has been severely restricted, however. 

 A sensitive probe of the presence of light gluinos is the running of
 $\alpha_s$\cite{ellisold}, as each  gluino changes the number of active
 flavours by 3. 
Here we show the sensitivity of the moments to the presence of a light
gluino with mass around 1 GeV. 
It is assumed  that the effect of the light gluino comes in only
through the running coupling and not through its effect on the final state
structure. This can be justified by noting that gluino pair production 
-- like quark pair production -- does
not contribute to the cascade evolution in leading double log order. 
In Fig.~\eref{momentsflavour} the lowest curve represents
 the predictions for the  moments assuming the presence 
of one light gluino with 1 GeV
 mass, i.e., 3  additional flavours at 
$\frac{\sqrt{s}}{2} \geq$  4 GeV.

The multiplicity $\N$ can be fitted again by readjusting the normalization 
parameters in agreement with previous findings \cite{cuypers}. 
On the other hand, the predictions for the higher moments,
especially with $q$ = 2 and $q$ = 3,  are far off the data. 
The energy dependence of the moments in presence of the light gluino is 
weaker in the same way as the running of $\alpha_s$ is weaker, as it is
expected from eq.~\eref{kq}. 
We conclude that the existence of a light gluino is not supported by our
analysis. However, it seems premature to definitely exclude such a particle at
present from this study. There are some simplifications in the present analysis
and our QCD fit without gluino is not perfect in a $\chi^2$ sense. 
Since the moments are very sensitive
to the existence of a light gluino, a meaningful statistical test could be performed 
after a further improvement of the theoretical description.

\subsection{Results for fixed $\alpha_s$} 
 
To see the effect of the running coupling in the inclusive energy
spectra, let us now consider for comparison the corresponding 
model with fixed coupling. 
Fig.~\eref{test3f} contains the same data as  Fig.~\eref{test3r}, but
the theoretical predictions refer now to the MLLA with fixed $\alpha_s$ 
\cite{lo}. The curves correspond   to 
three different values of the coupling $\alpha_s$, i.e., of the anomalous
dimension $\gamma_0$. For the multiplicity we take 
eq.~\eref{multtwopara} with  $\N_{part}$ given by eq.~\eref{norm:fix:full}. 
Again, the parameter $\gamma_0$ cannot be extracted from 
the study of the mean multiplicity alone. 
Including the other two cumulants, one can reproduce at best, choosing  
$\gamma_0$ = 0.64 (i.e., $\alpha_s$ = 0.21),  
the multiplicity data and the energy slope of 
the first moment $\xi_E$, though not its absolute value. 

In Fig.~\eref{figmoments} we show the predictions of the fixed-$\alpha_s$
model with $\gamma_0$ = 0.64 (dashed lines), where the absolute 
normalization is determined again at
threshold as in the case of the running coupling, see~\eref{xiqfix}. 
We also investigated whether the agreement with data can be improved if
the normalization at threshold is abandoned and shifted to a higher energy. 
To this end we
introduced an additional parameter for each cumulant, which allows for
vertical shifts of the curves; they have been chosen to fit 
the experimental points at $\sqrt{s}$ = 44 GeV.  In the following we will refer
to this model as the shifted fixed-$\alpha_s$ model. 

Contrary to the case of running $\alpha_s$, the vertical shifts can improve 
the description of the
moments for the model with fixed coupling, but only in a limited range of $cms$
energies. Especially, the
moments with $q\geq 3$ show a rather different trend with energy in
comparison with the data.

In an alternative investigation of 
the relevance of the running coupling for the inclusive energy spectra,
it has been proposed                                                                                                                                                                                                                                                                                                                                                                                   
to look at the predictions of Monte Carlo programs with and
without running coupling \cite{botterweck}. 
It was found that the  JETSET Monte Carlo \cite{jetset} with the
standard hadronization phase but the
coupling frozen at the value of $\alpha_s= 0.2$ in the perturbative phase 
describes the experimental data reasonably well throughout the PETRA/PEP
energy range and deviations occur only at higher energies.
At first sight this result seems to contradict our findings.
Note, however,
that in the JETSET Monte Carlo the perturbative evolution stops 
at a cut-off value
of about 1 GeV, when the string fragmentation takes over.
In our perturbative approach, we allow on the contrary the perturbative cascade
to evolve down to the much smaller cut-off
$Q_0\sim 270$ MeV. It is in this 
low energy domain that the variation of the coupling is most pronounced. 
The running coupling becomes large especially for small $k_\perp$ so that 
particles tend to be produced collimated. 
The perturbative calculations at low scales with running coupling 
seem  to simulate the production and decay of resonances 
implemented in Monte Carlo programs like JETSET. This supports the idea
that the parton
cascade with running coupling down to small scales is dual to the 
cascade with a shorter perturbative phase but 
with hadronic resonances in the last stage \cite{lo}.

\subsection{Rescaled cumulants}

In  addition to the standard moment analysis 
performed in the previous subsection, let us also consider 
the rescaled cumulants $K_q/\bar \xi$. 
These quantities become energy independent in case of fixed coupling 
 at high energies  as follows directly from \eref{kqlim}. In particular for the
 first three rescaled cumulants, one has
\begin{equation}
\frac{K_2}{\bar \xi} \simeq \frac{\gamma_0^2}{2 \bar \gamma_0^2} \frac{1}{\eta
+ \bar \gamma_0} \quad , \quad 
\frac{K_3}{\bar \xi} \simeq - 3 \frac{\gamma_0^2}{4 \bar \gamma_0^4} \frac{\eta}{\eta
+ \bar \gamma_0} \quad , \quad 
\frac{K_4}{\bar \xi} \simeq \frac{3 \gamma_0^2}{8 \bar \gamma_0^6} \frac{4 \eta^2 -
\gamma_0^2}{\eta + \bar \gamma_0}
\label{kqoverxi}
\end{equation}
Therefore these ratios
exhibit more directly the differences to the case of running coupling. 
In Fig.~\eref{reduced} the experimental data on these ratios, as derived from
our moment results in Table~1,  are compared
to the MLLA predictions with running $\alpha_s$, then with the 
fixed $\alpha_s$ and the shifted fixed-$\alpha_s$ models. 
Once again, a good description of data is given by the MLLA model with
running coupling. 
The fixed-$\alpha_s$ model shows the expected behaviour, i.e., the rescaled
cumulants tend to a constant value at large cms energies; its predictions lie
far away from the experimental data. 
The odd moments, which vanish in the DLA, approach their asymptotic limits
more slowly.
In the shifted fixed-$\alpha_s$ model previously described, 
the behaviour of the rescaled cumulants changes 
and the predictions become closer to the data. 
However, the deviations from the running $\alpha_s$ predictions become
obvious for the smaller and also the very high energies with $Y>6$, where the
ratios reflect the different asymptotic trends. For example, the ratio
$K_4/\bar\xi$ changes curvature when going from fixed to running
$\alpha_s$.
Data at these higher energies are becoming now available at 
the TEVATRON\cite{Korytov} and could give new information. 


\section{Analysis of the shape}

\subsection{Energy evolution of the shape}

The moment analysis has selected the solution with 
similar values for the parameters $Q_0$ and $\Lambda$; let us now consider the
predictions for the shape of the spectrum itself. 
Fig.~\eref{shift} shows the inclusive energy spectra $E dn/dp$ as
a function of  $\xi_E$, extracted from experimental 
data \cite{slac,topaz,tasso,opal,aleph15} using the fitted cut-off parameter 
$Q_0$ = 270 MeV as the effective mass in the calculation of the particle
energy. The curves show the predictions of the limiting spectrum with 
the same value of the $Q_0$ parameter.
The normalization has been fixed by choosing the integral 
of the spectrum to be equal to the average multiplicity 
according to the formula~\eref{multtwopara}, the respective numbers 
are  also  given in Table~1. 
The fit describes well the main features of the data  in the 
wide range of $cms $ energies $7\leq\sqrt{s}\leq 140$ GeV, especially in
the region with $\xi_E$ smaller than the peak position. 

Some deviations of the fitted curves from the
 data  can be seen for larger $\xi_E$'s, i.e., for 
 smaller particle energies. At low $cms$ energies the curves fall
somewhat above the data  near the peak position and below the data near
the kinematic limit ($\xi \to Y$). This behaviour may be related to the fact
that the limiting spectrum approaches a constant value and not zero 
for $\xi \to Y$, as expected for the exact solution of the evolution
equations (\ref{dqxiy}) or \eref{gln}. 

A iterative approximate solution of the MLLA equations which is valid 
 in the soft region and goes to zero for $\xi \to Y$ has been given in 
\cite{klo2}: 
\begin{equation}
D(\xi,Y,\lambda)_{MLLA} = D(\xi,Y,\lambda)|_{DLA} \exp \biggl[-a\int^Y_\xi
\frac{\gamma_0^2(y)}{4N_C} dy \biggr] 
\label{iterative}
\end{equation}
where 
\begin{equation}
D(\xi,Y,\lambda)|_{DLA} 
= \frac{4 C_A}{b}  \ln \left( 1 + \frac{Y-\xi}{\lambda}  \right) 
\left[ 1 + \frac{\frac{4N_C}{b} \int_0^{Y-\xi} 
d\tau \ln (1 + \frac{\tau}{\lambda}) 
\ln (1 + \frac{\xi}{\tau+\lambda})}{\ln (1 + \frac{Y-\xi}{\lambda})} \right] 
+ \dots
\label{duetermini}
\end{equation}
In Fig.~\eref{itera} the theoretical predictions from 
this approximation are 
compared with the same data as in Fig.~\eref{shift} for low particle energies 
$E<1$ GeV.     
A rather good description of data in the soft 
region  is obtained in this way, in agreement with previous findings
on the quantity $E dn/d^3p$ \cite{klo2}.

\subsection{High energy limit of the spectrum and $\zeta$-scaling} 


In QCD the asymptotic behaviour of the soft production phenomena can be
derived within the DLA, which takes into account the leading collinear and soft
singularities originating from the Bremsstrahlung processes. 
A well known scaling law of this type is the KNO scaling \cite{KNO,poly,bcm1}
of the particle multiplicity distribution for rescaled probabilities and
multiplicities.

Within QCD the momentum spectra do not scale asymptotically in the
Bjorken or Feynman variables $x=p/P$ because of the Bremsstrahlung emissions 
with large transverse momenta, as is well known. 
Instead, they approach a finite scaling limit in certain rescaled logarithmic
variables. This proposal, put forward already more than ten years ago
\cite{dfk}, has never been studied since. A scaling behaviour 
of similar type for angular
correlations  has been recently proposed \cite{ow} and found some support by
experimental data \cite{bm}.

The asymptotic behaviour of the inclusive energy spectrum 
can be derived from the DLA evolution equation, i.e. \eref{gln} with $a=0$.
One finds a scaling law in the rescaled logarithmic variable $\zeta$
for the rescaled spectrum \cite{dfk}   
\begin{equation}
\frac{\ln D(\xi,Y)}{\ln \N (Y)} = F(\zeta) \quad , \quad \zeta = \frac{\xi}{Y} 
\label{zetascal}
\end{equation}
independent of $Y$. 
The logarithmic variables naturally occur in the asymptotic limit
as they absorb the Bremsstrahlung singularities in 
the evolution equation~\eref{Devolve}.

The scaling function $F(\zeta)$ takes in DLA the following form~\cite{dfk}: 
\begin{equation}
F(\zeta) = \frac{\mu}{\sinh \mu} \qquad \hbox{with} \qquad 2 \zeta - 1 =
\frac{\sinh 2\mu - 2\mu}{2 \sinh^2 \mu} 
\label{zetadla}
\end{equation} 
The function $F(\zeta)$ is symmetric around $\zeta$ = 1/2, where it has a 
maximum of 1, and it goes as $\sqrt{\zeta \ln 1/\zeta}$ for $\zeta \to 0$. 
In the MLLA model with fixed coupling  
explicit analytical expressions for the spectrum and for the
average multiplicity are available\cite{lo} 
and the scaling function is given by $F(\zeta) = 2 \sqrt{\zeta (1 - \zeta)}$. 

For further illustration of the scaling law~\eref{zetascal}, 
we recall the Gaussian approximation of the DLA~\cite{DFK,MUEL} 
\begin{equation}
D(\xi,Y) \simeq \frac{\N(Y)}
{((Y+\lambda)^{3/2}-\lambda^{3/2})^{\frac{1}{2}}}
\exp\left(-\frac{3\sqrt{\frac{4N_C}{b}} (\xi-\frac{Y}{2})^2}
                        {(Y+\lambda)^{3/2}-\lambda^{3/2}}\right)
\label{Dgauss}
\end{equation}
As the multiplicity $\N(Y) \sim \exp ( \sqrt{16 N_C (Y+\lambda)/b})$ 
at high energies, one finds in exponential accuracy 
\begin{equation}
F(\zeta) \simeq  1-\frac{3}{2}(\zeta-\frac{1}{2})^2
\label{Fgauss}
\end{equation} 
which depends only on $\zeta$ and not on $Y$. In case of fixed $\alpha_s$ 
the Gaussian is slightly narrower $F(\zeta) \simeq 1-2(\zeta-\frac{1}{2})^2$. 
The DLA asymptotic scaling functions both for running and fixed $\alpha_s$ 
and their Gaussian approximations are shown in Fig.~\eref{zetatheory}. 
We notice that the two  asymptotic curves for running and fixed $\alpha_s$ 
do not differ very much 
quantitatively in the region around the maximum $\zeta \sim 1/2$, but they 
do in the soft region, where the effect of the running of $\alpha_s$ becomes
relevant~\cite{klo2} and the corresponding curve shows indeed a steeper slope
than the fixed-$\alpha_s$ one.  
Notice also that the Gaussian approximation considerably deteriorates towards
the limits $\zeta \sim 0$ and $\zeta \sim 1$. 

It should be noted that only the transformed observable~\eref{zetascal} 
and not the spectrum $D(\zeta,Y)$ itself approaches a finite limit. The
spectrum $D(\zeta, Y)$ can be approximated by a Gaussian in $\zeta$ with width
$\sigma_{\zeta}^2 \sim 1/\sqrt{Y}$ and therefore one finds asymptotically for
$D$ and its moments 
\begin{equation} 
D(\zeta,Y) \to \delta(\zeta -\frac{1}{2}) \quad , \quad <\zeta^q> \to
\frac{1}{2^q} 
\end{equation}  
The MLLA results for the moments~\eref{lavg:LS}--\eref{l4:LS} 
indeed approach this limit. 

In Fig.~\eref{zeta} we show the experimental data for the observable 
\eref{zetascal} at three different $cms$ energies \cite{tasso,topaz,opal15}. 
In this figure $\xi$ is chosen  to
be $\xi_p$ and $Y=\ln(\sqrt{s}/2Q_0)$ with  $Q_0$ 
= 270 MeV. Experimental data for the average charged 
multiplicity have been taken from~\cite{tassomult,topazmult,opal15}. 
Fig.~\eref{zetae} shows the same observable, but this time as a function of 
$\xi$ = $\xi_E$, where the energy is calculated with 
the effective mass $Q_0$ = 270 MeV. In this way, experimental spectra have a
common kinematical boundary at $\zeta_E = 1$ and can be compared with the
theoretical predictions as in Fig.~\eref{shift}. 
Fig.~\eref{zetae} also shows the theoretical predictions of the Limiting 
Spectrum normalized by the  average multiplicity~\eref{multtwopara} and 
~\eref{norm:LS} at the same
$cms$ energies of the experimental data, as well as the asymptotic DLA prediction of
eq.~\eref{zetadla}.  

One can see from these figures that the original energy evolution of the
spectrum visible  in 
Fig.~\eref{shift} is largely removed if the scaling variables 
\eref{zetascal} are used. Some scaling violation remains, 
especially in the small $\zeta$ region.
The limiting spectrum  reproduces well the small scaling violations 
shown by the data. 
It approaches the asymptotic limit very slowly and only at unphysically large 
energies\footnote{For instance the scaling 
function $F(\zeta)$ at $\zeta$ = 1/2, whose asymptotic value is 1,  
reaches 0.8 at $Y \sim$ 68 ($\sqrt{s} \sim  10^{28}$ GeV) and 
0.9 at $Y \sim$ 460 ($\sqrt{s} \sim  10^{200}$ GeV).}.  
The data at the present $cms$ energies lie far away from 
the asymptotic DLA curve in the small $\zeta$ domain; however, it is interesting
to notice that  the data for soft particles ($p \sim 
0$, $\xi \sim Y$, $\zeta \sim 1$) are already close to the asymptotic limit. 
This result lends further support to the idea that very soft
particles are basically determined by the nearby Bremsstrahlung singularities
and not affected by the non singular terms in the splitting
functions taken into account in MLLA 
nor by the recoil effects (see \cite{klo2} for a previous 
discussion of this point). 

The energy dependence of the position of the maximum of the $\zeta$
distribution, $\zeta^*$,  
is shown in Fig.~\eref{zetamax} up to LEP-1.5 energies. 
Also in this case, 
the  data  closely follow the prediction of the limiting spectrum, 
which very slowly approaches the asymptotic DLA limit 
$\zeta^*=\frac{1}{2}$.

One concludes that the data show an approximate scaling law
in the presently available energy range.
However, this 
scaling is preasymptotic, i.e., the asymptotic shape of the distribution is
quite different from the one observed at present energies.
Similar results on the existence of a large preasymptotic scaling 
regime have also been predicted for the multiplicity scaling (KNO)
and its violation \cite{mw,yld}; in this case significant deviations
from scaling are expected only around $\sqrt{s}\sim 1$ TeV \cite{ugl}.

\section{Conclusions} 

The perturbative QCD approach has been shown to describe well 
experimental data on  charged particle inclusive energy spectra 
in $e^+e^-$ annihilation. The new features of the moment analysis 
 of the spectrum have been discussed. 
 
 The first determination of
 the two independent essential parameters of the theory has been performed, whereby 
the best description of the data 
is obtained for $Q_0 \simeq \Lambda \simeq $ 270 MeV with an uncertainty of
about 20 MeV. 
 The dependence of the moments on the initial conditions has been studied; 
 if one takes the nonperturbative initial condition of the perturbative QCD
 evolution at the threshold of the process, a good description of 
 the moments also in their absolute normalization is obtained in the full  
 energy range available including the low energies of a few GeV. 
 The sensitivity of the moments to the running of the coupling has been
 established by comparing the predictions of the full model with 
 the predictions of a model with frozen coupling. 
 The effect of heavy quark thresholds 
 in the running of the coupling has also been discussed; 
a good  phenomenological description of the data
at present energies is obtained already by including only 
three active flavours.
The moments are very sensitive to the presence of a light gluino with mass
around 1 GeV, but there is no evidence for the type of effect expected. 

  The inclusive energy spectra themselves have also been studied. 
The limiting spectrum with $Q_0 = \Lambda$ 
is found to provide a good overall description of the
data in a large $cms$ energy range. An improvement in the soft region 
can be obtained by applying an approximation which takes 
into account the boundary conditions at $\xi  = Y$ explicitly.

Data in the presently available $cms$ energy range 
support an approximate scaling law, the $\zeta$-scaling, 
predicted at asymptotic energies. Some violations of the scaling
behaviour in this preasymptotic energy range are observed, as expected from
the MLLA. The data in the soft energy region ($\zeta \lsim 1$) at available
energies are already close to the asymptotic predictions.

There are some simplifications both in the phenomenological analysis (for
example, neglect of difference in light and heavy quark fragmentation) and the
theoretical description. The respective improvements would yield a more
stringent test of the theoretical scheme and possibly a quantitative improvement
of the fit to the data. 
As the main result of the present analysis we consider 
the good description  of the data down to low scales -- even
if not entirely quantitatively everywhere -- in
agreement with QCD with running $\alpha_s$ and the LPHD picture.

\newpage 

\appendix
\section{Explicit Formulae for Moments}

\subsection{Moments in the general case}

The functions entering eq. (\ref{moments:full}) are defined as
follows\cite{DKTInt}: 
$\N={\mathcal N_1}+{\mathcal N_2}$ is composed of two terms
increasing and decreasing with energy, respectively,
\begin{gather}
{\mathcal N_1} =  \Gamma(B) \Gamma(1-B) z_1 \left( \frac{z_2}{z_1} \right)^B 
I_{B+1}(z_1) I_{-B}(z_2)\label{n1}\\
{\mathcal N_2} =  \Gamma(-B) \Gamma(B+1) z_1 \left( \frac{z_2}{z_1} \right)^B 
I_{-B-1}(z_1) I_{B}(z_2)\label{n2}\\
z_1 = \sqrt{ 16 N_c (Y + \lambda)/b  } ; \qquad
z_2 = \sqrt{ 16 N_c\lambda/b }.
\nonumber
\end{gather}
The two terms build up the total multiplicity 
%
%
\begin{equation}
\N = z_1 \left( \frac{z_2}{z_1} \right)^B \left[ 
I_{B+1}(z_1) K_B(z_2) + K_{B+1}(z) I_B(z_2) \right] 
\label{mllamult}
\end{equation}
The functions $ L_k^{(q)}$ and $R_k^{(q)}$ in (\ref{moments:full})
are computed from
\begin{align}
L_k^{(q)} &= D_{q-k}(B+1,B+2,z_1)D_k(-B,1-B,z_2) \label{lkq}\\ 
R_k^{(q)} &= D_{q-k}(0,-B,z_1)D_k(0,B+1,z_2) \label{rkq}\\
D_k(g,c,z)& = P_0^{(q)}(g,c,z) + \frac{2}{z}
\frac{I_{c}(z)}{I_{c-1}(z)} P_1^{(q)}(g,c,z) \nonumber
\end{align}
where $P_i^{(q)}$ are the polynomials
\begin{equation}
P_0^{(q)}(g,c,z) = \sum_{k=0}^{q-1} \alpha^{(q)}_{q-k} \left( \frac{2}{z}
\right)^{2k}, \qquad  
P_1^{(q)}(g,c,z) = \sum_{k=0}^{q-1} \beta^{(q)}_{q-k} \left( \frac{2}{z}
\right)^{2k}. \label{p01}
\end{equation}
The coefficients of highest order are given by
\begin{equation}
\alpha^{(q)}_q = \frac{1}{2^q}, \qquad \beta^{(q)}_q = \frac{q}{2^q} \left( 
B + \frac{q-1}{3} \right) 
\end{equation}
whereas in lower order
they can be found by solving a 
$q\times q$ linear  system of equations. For the first four moments 
one finds explicitly ($(a)_n \equiv \Gamma(a+n)/\Gamma(a) = a(a+1) \dots
(a+n-1)$): 
\begin{align}
\beta^{(q)}_1 &= \frac{\Phi^{(q)}_{1-c}}{1-c} , \quad 
\beta^{(q)}_2 = \frac{\Phi^{(q)}_{2-c}}{(c-2)_2} - 
\frac{\Phi^{(q)}_{1-c}}{(c-1)_2} \nonumber \\ 
\beta_3^{(q)} &= - \frac{1}{2} \frac{\Phi^{(q)}_{3-c}}{(c-3)_3} + 
\frac{\Phi^{(q)}_{2-c}}{(c-2)_3} - \frac{1}{2}
\frac{\Phi^{(q)}_{1-c}}{(c-1)_3} 
\end{align}

\begin{align}
\alpha^{(q)}_1 &= \frac{\Phi^{(q)}_{1-c}}{(c-1)_2} , \quad 
\alpha^{(q)}_2 = - \frac{\Phi^{(q)}_{2-c}}{(c-2)_3} - 
\frac{\Phi^{(q)}_{1-c}}{(c-1)_3} \nonumber \\ 
\alpha_3^{(q)} &= \frac{1}{2} \frac{\Phi^{(q)}_{3-c}}{(c-3)_4} -  
\frac{\Phi^{(q)}_{2-c}}{(c-2)_4} + \frac{1}{2}
\frac{\Phi^{(q)}_{1-c}}{(c-1)_4} 
\end{align}
in terms of the expressions
\begin{equation}
\Phi^{(1)} = \frac{1}{2} (n-1)_2 + g n 
\end{equation}
\begin{equation}
\Phi^{(2)} = \frac{1}{4} (n-2)_3 + ( g + \frac{2}{3} ) (n-2)_3 
+ (g)_2  (n-1)_2 
\end{equation}
\begin{eqnarray}
 \Phi^{(3)} &=& \frac{1}{8} (n-5)_6 
 + (\frac{3}{4} g + 1) (n-4)_5  \nonumber \\
 &+& ( \frac{3}{2} g^2 + \frac{7}{2} g + \frac{3}{2}) (n-3)_4 
 + (g)_3 (n-2)_3
\end{eqnarray}
\begin{eqnarray}
\Phi^{(4)} &=& \frac{1}{16} (n-7)_8  + (\frac{1}{2} g + 1) (n-6)_7  
 +  ( \frac{3}{2} g^2 + \frac{11}{2} g + 
 \frac{13}{3}) (n-5)_6  \nonumber \\ 
 &+& (2 g^3 + 10 g^2 + 14 g 
 + \frac{24}{5}) (n-4)_5 + (n-3)_4. 
\end{eqnarray}

\subsection{Limiting spectrum}
In the special case $Q_0=\Lambda$ we find~\cite{DKTInt} for
the moments $q \le 4$ explicitly:
\begin{equation}
\frac{\bar \xi}{Y} = \frac{1}{2} + \frac{B}{z}
\frac{I_{B+2}(z)}{I_{B+1}(z)}
\label{lavg:LS}
\end{equation}
\begin{equation}
\frac{< \xi^2 >}{Y^2} = \frac{1}{4} +   \frac{B(B+\frac{1}{3})}{z^2}
+ \frac{(B+ \frac{1}{3})}{z} \left( 1 - \frac{2 B (B+2)}{z^2} \right)
\frac{I_{B+2}(z)}{I_{B+1}(z)}
\label{l2:LS}
\end{equation}
\begin{eqnarray}
\label{l3:LS}
\frac{< \xi^3 >}{Y^3} &=& \frac{1}{8} +   \frac{3 B(B+1)}{2 z^2}
- \frac{ 2 B^2 (B+1) (B+3)}{z^4}   +  \\
&+& \frac{2}{z} \left[
\frac{3B+2}{8} -  \frac{B (B+1)(B+3)}{z^2}
+ \frac{4 B (B)_4}{z^4}  \right]
\frac{I_{B+2}(z)}{I_{B+1}(z)} \nonumber
\end{eqnarray}
and
\begin{eqnarray}
\label{l4:LS}
\frac{< \xi^4 >}{Y^4} &=& \frac{1}{16} + \alpha_3^{(4)}
\left( \frac{2}{z} \right)^2
+ \alpha_2^{(4)} \left( \frac{2}{z} \right)^4
+ \alpha_1^{(4)} \left( \frac{2}{z} \right)^6  + \\
&+& \frac{2}{z} \left[
\frac{B+1}{4}
+ \beta_3^{(4)} \left( \frac{2}{z} \right)^2
+ \beta_2^{(4)} \left( \frac{2}{z} \right)^4
+ \beta_1^{(4)} \left( \frac{2}{z} \right)^6  \right]
\frac{I_{B+2}(z)}{I_{B+1}(z)} \nonumber
\end{eqnarray}
where  
\begin{equation}
\alpha_3^{(4)} =  10 \frac{\Phi_{-B}^{(4)}}{(B)_5} -
\frac{1}{4} \frac{B(B+1)(B+3)}{B-1} 
\end{equation}
\begin{equation}
\alpha_2^{(4)} = - 3 \frac{\Phi_{-B}^{(4)}}{(B)_4}
\quad , \quad
\alpha_1^{(4)} =  \frac{\Phi_{-B}^{(4)}}{(B+1)_2} \nonumber 
\end{equation}
and
\begin{equation}
\beta_3^{(4)} =  - 6 \frac{\Phi_{-B}^{(4)}}{(B-1)_5} +
\frac{1}{4} \frac{(B)_4}{B-1}
\end{equation}
\begin{equation}
\beta_2^{(4)} = 2 \frac{\Phi_{-B}^{(4)}}{(B)_3}
\quad , \quad
\beta_1^{(4)} = - \frac{\Phi_{-B}^{(4)}}{(B+1)} \nonumber 
\end{equation}
and
\begin{align}
\Phi_{-B}^{(4)} &= \frac{1}{16} B^8 + \frac{3}{4} B^7 + 3.45833 B^6 +
7.7 B^5 \\ 
&+ 8.39583 B^4 + \frac{15}{4} B^3 + \frac{1}{12} B^2 - \frac{1}{5} B. \nonumber
\end{align}

\subsection{Fixed coupling}

The first coefficients $A_q,~B_q$  
in (\ref{xiqfix}) read (here $\rho \equiv \eta/\bar \gamma_0$)
\begin{align}
A_1 &= (\rho^2 + 1) \frac{Y}{2} 
\quad , \quad B_1 = (- \rho^2 - 1) \frac{1}{2\bar \gamma_0} + \rho Y \; ;  \\ 
A_2 &= (- 3 \rho^3 - \rho) \frac{Y}{4 \bar \gamma_0} + (3 \rho^2 + 1)
\frac{Y^2}{4}  \; , \\ 
B_2 &= (3 \rho^3 + \rho) \frac{1}{4 \bar \gamma_0^2} + (-3 \rho^2 -1 ) 
\frac{Y}{4\bar \gamma_0} + (\rho^3 + 3 \rho) \frac{Y^2}{4} \; ;\nonumber 
\end{align} 
\begin{align}
A_3 &= (15 \rho^4 - 3) \frac{Y}{8 \bar \gamma_0^2} - 12 \rho^3 
\frac{Y^2}{8\bar \gamma_0} + (\rho^4 + 6 \rho^2 + 1) \frac{Y^3}{8} \; , \\ 
B_3 &= (-15 \rho^4 + 3) \frac{1}{8 \bar \gamma_0^3} + 12 \rho^3
\frac{Y}{8\bar \gamma_0^2} + (- 6 \rho^4 - 6 \rho^2) \frac{Y^2}{8 \bar
\gamma_0} + (4 \rho^3 + 4 \rho) \frac{Y^3}{8} \; ; \nonumber \\
A_4 &=  (- 105 \rho^5 + 30 \rho^3 + 27 \rho) \frac{Y}{16 \bar \gamma_0^3} 
+ (75 \rho^4 - 18 \rho^2 - 9) \frac{Y^2}{16 \bar \gamma_0^2}  \\ 
& \qquad + 2 (-5 \rho^5 - 14 \rho^3 + 3 \rho) \frac{Y^3}{16 \bar \gamma_0} 
+ (5 \rho^4 + 10 \rho^2 + 1) \frac{Y^4}{16} \; , \nonumber \\ 
B_4 &= (105 \rho^5 - 30 \rho^3 - 27 \rho) \frac{1}{16 \bar \gamma_0^4}  
+ (-75 \rho^4 + 18 \rho^2 + 9) \frac{Y}{16 \bar \gamma_0^3} \nonumber \\ 
& \qquad + (45 \rho^5 + 18 \rho^3 - 15 \rho) \frac{Y^2}{16 \bar \gamma_0^2} 
+ 2 (-15 \rho^4 - 2 \rho^2 + 1) \frac{Y^3}{16 \bar \gamma_0} \nonumber \\ 
& \qquad  + (\rho^5 + 10 \rho^3 + 5 \rho) \frac{Y^4}{16} \; . \nonumber
\end{align} 

\newpage

\section*{Table Caption}

{\bf Tab. 1}:
The average multiplicity $\N_E$, the average
value $\bar \xi_E$, the dispersion $\sigma^2$, the skewness $s$ and the
kurtosis $k$ of charged particle inclusive 
energy spectra $E dn/dp$ vs. $\xi_E$ for 
$Q_0$ = 270 MeV at various $cms$ energies
$\sqrt{s}$. 
In brackets theoretical predictions of the limiting spectrum of MLLA with
running $\alpha_s$; the second entry in the average multiplicity
column contains the results of eq.~\eref{multtwopara} 
with eq.~\eref{norm:LS}; the first one the results with $c_2 = 0$. 
Errors on the average multiplicity data points 
include both statistical and systematic errors. 
Results at LEP and LEP-1.5 $cms$ energies from~\cite{klo1}.

\newpage 

\section*{Figure Captions}

{\bf Fig. 1: a)} Comparison of 
charged particle inclusive single particle spectra, $E dn/dp$ vs. $\xi$,  
at $\sqrt{s}$ = 14  GeV\cite{tasso} for different mass  assigment: 
the inclusive momentum spectrum $p dn/dp$ vs. $\xi_p$ (diamonds) and the 
rescaled spectra 
$E dn/dp$ vs. $\xi_E$, with $E^2 = p^2 + Q_0^2$, $Q_0$ = 138 MeV
(triangles) and $Q_0$ = 270 MeV (squares). Also shown are the upper limits 
of $\xi_E$  given by $Y =
\ln (\sqrt{s}/2Q_0)$; {\bf b)}: same as in {\bf a)}, but 
at $\sqrt{s}$ = 91  GeV\cite{opal}.

\medskip\noindent 
{\bf Fig. 2}: 
Dependence on $Y$ of the three lowest order  
moments of the inclusive energy spectrum. 
The data 
of the mean multiplicity $\N_E$, average value $\xi_E$ and dispersion
$\sigma^2$ of inclusive energy spectra
for $Q_0$ = 138 MeV (triangles), $Q_0$ = 270 MeV (diamonds) and $Q_0$ = 
350 MeV (squares) are compared 
with theoretical predictions of MLLA with running coupling 
with $\lambda$ = 0 (limiting spectrum)
(solid line), $\lambda$ = 0.5 (dashed line), $\lambda$ = 4 (dotted line). 
The theoretical predictions for the mean multiplicity are computed from 
eqs.~\eref{multtwopara} and~\eref{norm:LS}. 

\medskip\noindent 
{\bf Fig. 3}: 
Same data as in Fig. 2 for different $Q_0$ parameters;  
comparison with theoretical predictions of MLLA with
fixed coupling with $\gamma_0$ = 0.64 (solid line), $\gamma_0$ = 
0.4 (dashed line), $\gamma_0$ = 1 (dotted line). 
The theoretical predictions for the mean multiplicity are computed from 
eqs.~\eref{multtwopara} and ~\eref{norm:fix:full}. 

\medskip\noindent 
{\bf Fig. 4}: 
The average multiplicity $\N_E$ and the first four 
cumulants of charged particle energy spectra $E dn/dp$ vs. $\xi_E$, 
are shown as a function of $Y = \ln (\sqrt{s}/2Q_0)$ for $Q_0$ = 270 MeV 
at various $cms$ energies 
(see  Table~1). 
The curves show the predictions of the limiting
spectrum with $Q_0$ = 270 MeV with 3 active flavours (solid line), the number
of flavours $n_f$ variable with the heavy quark thresholds at $\sqrt{s}/2 
= 4 m_Q$ (dashed
line) and the inclusion of a light gluino with mass of 1 GeV (dotted line). 
In all cases,  the mean multiplicity is computed from 
eqs.~\eref{multtwopara} and ~\eref{norm:LS}.

\medskip\noindent 
{\bf Fig. 5}: 
Same data as in Fig.~4, but the curves show the 
predictions of the limiting spectrum (i.e. $Q_0 = \Lambda$) 
of MLLA with running $\alpha_s$ (solid line),  
 of MLLA with fixed $\alpha_s$ (dashed line) 
 and of the shifted  fixed $\alpha_s$ model (dotted line); $n_f$ = 3 
 everywhere.

\medskip\noindent 
{\bf Fig. 6}: 
Rescaled cumulants $K_q/\bar \xi_E$ 
as a function of $Y = \ln (\sqrt{s}/2 Q_0)$ for $Q_0$ = 270 MeV. 
Data as in Fig.~4 are compared with the corresponding 
predictions of the limiting spectrum of MLLA
with running $\alpha_s$ (solid line). 
The dashed curve shows the predictions of the model with
fixed $\alpha_s$ ($\gamma_0$ = 0.64); in this case, the rescaled cumulants 
approach constant values at high energies (0.534, 0.344, 0.528 respectively,
with the chosen value of $\gamma_0$). Predictions of the shifted  
fixed $\alpha_s$ model are also shown (dotted line). 

\medskip\noindent 
{\bf Fig. 7}: 
Comparison of inclusive spectra $E dn/dp$ vs. $\xi_E$, with
$E^2 = p^2 + Q_0^2$, $Q_0$ = 270 MeV at various $cms$ energies 
with predictions of  the limiting spectrum ($K_h$ fixed at each energy from the
fit of eqs.~\eref{multtwopara} and~\eref{norm:LS}). Each curve is shifted up by 0.5 for
clarity.

\medskip\noindent 
{\bf Fig. 8}: 
Comparison of inclusive spectra $E dn/dp$ vs. $\xi_E$, with
$E^2 = p^2 + Q_0^2$, $Q_0$ = 270 MeV at various $cms$ energies 
with predictions of the MLLA iterative solution~\eref{iterative} 
($Q_0$ = 270 MeV, $\lambda$ = 0.01, $K_h$ = 0.45). 
Data and predictions for particle energies $E \le$ 1 GeV  are shown. 

\medskip\noindent 
{\bf Fig. 9}: 
The DLA asymptotic prediction~\eref{zetadla} for the scaling function 
$F(\zeta)$, the prediction of the Gaussian Approximation~\eref{Fgauss} 
and the corresponding predictions with fixed $\alpha_s$ in comparison.  

\medskip\noindent 
{\bf Fig. 10}: 
Test of $\zeta$-scaling for the momentum spectra at 14, 58 and 130
GeV\cite{tasso,topaz,opal15} with average charged multiplicities taken
from \cite{tassomult,topazmult,opal15}. 

\medskip\noindent 
{\bf Fig. 11}: 
Test of $\zeta$-scaling for the energy spectra. Data as in Fig.~10, but 
the particle energy is calculated using the mass $Q_0$ = 270 MeV, in 
 comparison 
with theoretical predictions from the limiting spectrum ($Q_0$ = 270 MeV) 
normalized to the predicted average multiplicity 
according to eqs.~\eref{multtwopara} and~\eref{norm:LS}.  
The DLA prediction~\eref{zetadla} at  asymptotic $cms$ energy  is also shown.

\medskip\noindent 
{\bf Fig. 12}: 
Maximum of the rescaled inclusive momentum distribution 
$\zeta^*=\xi^*/Y$  
as a function of $Y = \ln \frac{\sqrt{s}}{2 Q_0}$; 
comparison between experimental data at various $cms$ 
energies\cite{tasso,opal,aleph15,delphi15,opal15,slac,slac2,topaz,aleph,delphi,l3} 
  and theoretical prediction in MLLA,  
numerically extracted from the shape of the limiting spectrum (solid line) 
for the cut-off parameter $Q_0=\Lambda$ = 270 MeV~\cite{klo1}. 
Crosses mark the predictions at the $cms$ energies 
200 GeV and 500 GeV. The asymptotical DLA result $\zeta^* =
\frac{1}{2}$ is also shown (dashed line).


{ \baselineskip=45pt

\begin{table}     
\vspace{-4.0cm}
\centerline{Table 1}
\vspace{1.0cm}
 \begin{center}
 \vspace{4mm}
 \begin{tabular}{||c|c|c|c|c|c||}
  \hline
 \vspace{-0.4cm}
 Exp. & & & & &  \\ 
 \vspace{-0.4cm}
  & $\N_E$ & $\bar \xi_E$ & $\sigma^2$ & $s$ & $k$ \\  
$\sqrt{s}$   (GeV)       &  &  &  & &  \\  
  \hline
 MARK I\cite{slac}  & 
 3.01$\pm$0.3 & 
 1.02$\pm$0.02 &  
 0.14$\pm$0.01 & 
 -0.50$\pm$0.17 & 
 -0.46$\pm$0.13   \\ 
 3.0 & (2.10,input) & (1.14)  & (0.13)   & (-0.50) & (-0.50) \\
 MARK I\cite{slac}  & 
 3.66\pp 0.37 & 
 1.27\pp 0.02 & 
 0.17\pp 0.01 & 
 -0.58\pp 0.08 & 
 -0.12\pp 0.20 \\ 
  4.03 &  (2.73,3.61) & (1.33) & (0.18) & (-0.48) &  (-0.51) \\ 
 MARK II\cite{slac2}  & 
     \pp  & 
 1.47\pp 0.04 & 
 0.19\pp 0.01 & 
 -0.57\pp 0.08 & 
 -0.20\pp 0.17 \\ 
  5.2 &  (3.36,4.21) & (1.49) & (0.22) & (-0.47) &  (-0.52) \\ 
 MARK I\cite{slac}  & 
 4.63\pp 0.46 &  
 1.61\pp 0.03 & 
 0.26\pp 0.01 & 
 -0.56\pp 0.08 & 
 -0.29\pp 0.18 \\ 
   7.4 &  (4.11,4.92) & (1.66) & (0.28) & (-0.45) &  (-0.52) \\ 
 TASSO\cite{tasso} & 
  7.64\pp 0.59   & 
 2.10\pp 0.04 & 
 0.40\pp 0.01 &  
 -0.47\pp 0.07 &  
 -0.42\pp 0.17 \\ 
   14. &  (6.73,7.40) & (2.12) & (0.44) & (-0.42) & (-0.52)  \\
 TASSO\cite{tasso} & 
  9.65\pp 0.68  & 
 2.38\pp 0.04 & 
 0.54\pp 0.02 & 
 -0.47\pp 0.07 & 
 -0.44\pp 0.18  \\ 
   22. & (8.87,9.44) & (2.40)  & (0.56)   & (-0.40)  & (-0.52) \\
 TASSO\cite{tasso} & 
   12.21\pp 0.86  & 
 2.67\pp 0.03 & 
 0.68\pp 0.01 & 
 -0.44\pp 0.03 & 
 -0.51\pp 0.09 \\ 
  35. & (11.51,11.95) & (2.69) & (0.70) & (-0.38)  & (-0.52)  \\
 TASSO\cite{tasso} & 
 13.38\pp 1.05  & 
 2.80\pp 0.03 & 
 0.75\pp 0.01 & 
 -0.40\pp 0.04 & 
 -0.59\pp 0.09 \\ 
 44. &  (12.96,13.33) & (2.82) & (0.76)  & (-0.37)  &  (-0.52) \\
 TOPAZ\cite{topaz} & 14.54\pp 0.43 & 
 3.01\pp 0.03 & 
 0.80\pp 0.02 & 
 -0.43\pp 0.05& 
 -0.49\pp 0.15 \\ 
 58. & (15.09,15.34) & (3.00) & (0.85) &  (-0.36) & (-0.52) \\ 
\hline 
ALEPH\cite{aleph}  & 
 18.81\pp 1.05 & 
 3.24\pp 0.04 &
 0.99\pp 0.05 & 
 -0.39\pp 0.10 & 
 -0.59\pp 0.32 \\
 DELPHI\cite{delphi} & 
 19.17\pp 1.00 & 
 3.32\pp 0.02 &
 1.03\pp 0.01 & 
 -0.40\pp 0.02 &
 -0.59\pp 0.07 \\
 L3\cite{l3}    & 
 18.74\pp 1.09 & 
 3.28\pp 0.06 & 
 0.99\pp 0.06 & 
 -0.35\pp 0.13 &
 -0.65\pp 0.40 \\
 OPAL\cite{opal}  &  
 18.95\pp 1.00 & 
 3.29\pp 0.01 & 
 0.99\pp 0.01 & 
 -0.36\pp 0.03 & 
 -0.59\pp 0.09 \\ 
 LEP-1 (avg)         &  
 18.93\pp 0.52  & 
 3.29\pp 0.01 & 
 1.01\pp 0.02 & 
 -0.39\pp 0.02 & 
 -0.59\pp 0.05 \\
 DELPHI--$\gamma$~\cite{delphi15} & 
 19.20\pp 0.26 & 
 3.33\pp 0.04 &
 0.99\pp 0.03 & 
 -0.46\pp 0.05 &
 -0.44\pp 0.18 \\  
  91.2 &  (input,input) & (3.27)  & (1.00)  & (-0.35) &   (-0.52)  \\
\hline 
 ALEPH~\cite{aleph15}   &  
   22.04\pp 0.47   & 
   3.52\pp 0.06  & 
   1.19\pp 0.04 & 
    -0.37\pp 0.07  & 
    -0.62\pp 0.26  \\
 DELPHI~\cite{delphi15}   & 
 22.27\pp 0.58  & 
 3.47\pp 0.05 & 
 1.13\pp 0.05 & 
 -0.40\pp 0.08 & 
 -0.49\pp 0.29 \\
 OPAL~\cite{opal15} &  
 21.50\pp 0.57 & 
 3.51\pp 0.07 & 
 1.19\pp 0.04 & 
 -0.35\pp 0.07 & 
 -0.63\pp 0.25 \\ 
 LEP-1.5 (avg)         &  
 21.95\pp 0.31  & 
 3.495\pp 0.034 & 
 1.185\pp 0.025 & 
 -0.365\pp 0.042 & 
 -0.59\pp 0.15 \\
133 &  (22.6,22.5) & (3.50)  & (1.14)  & (-0.34) &   (-0.52)  \\
  \hline
  \hline
 \end{tabular}
 \end{center}
\label{table}
\end{table} 

 \newpage

\begin{figure}[p]
\begin{center} 
\mbox{\epsfig{file=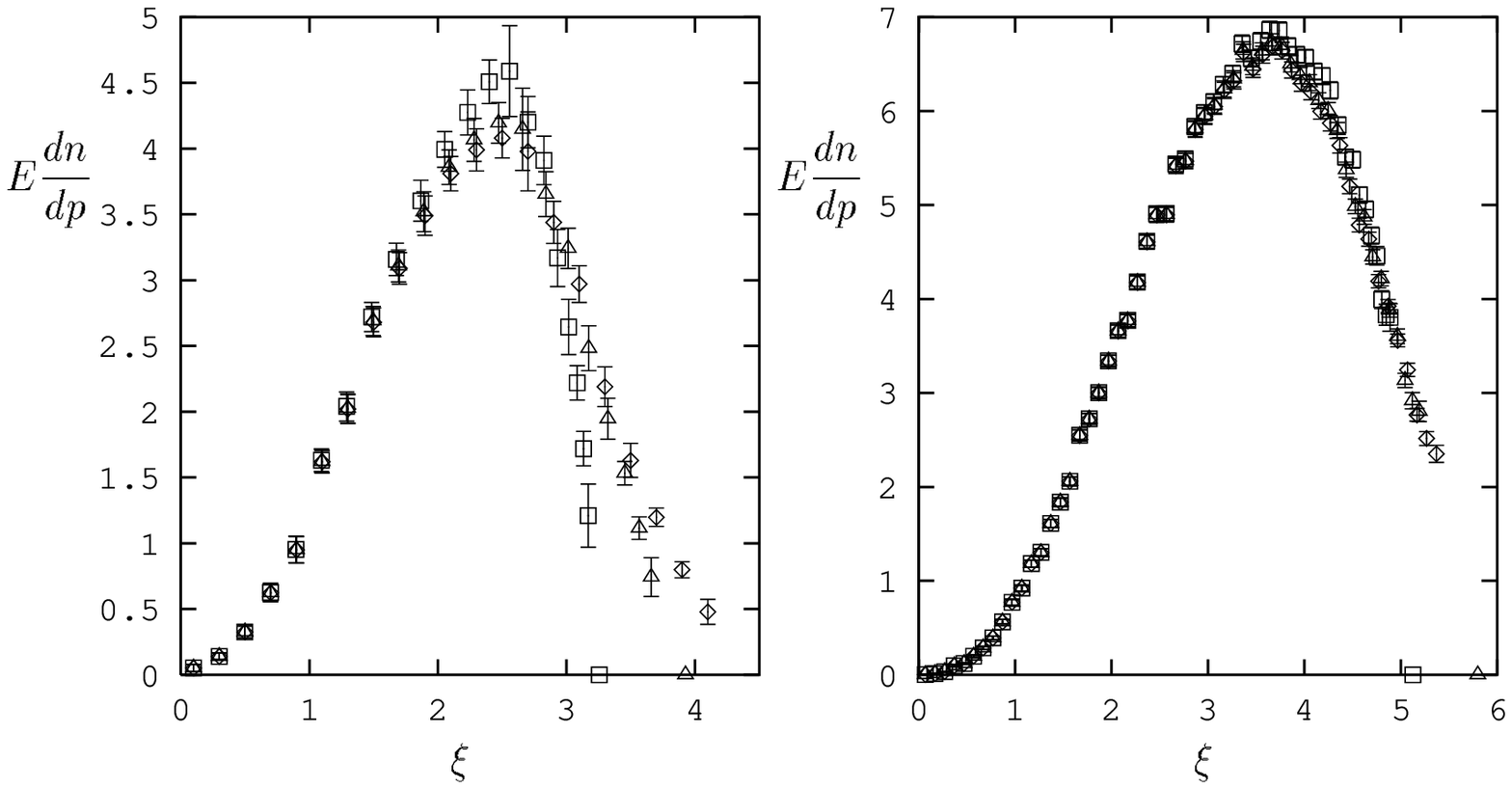,width=17cm,bbllx=4.cm,bblly=6.cm,bburx=21.cm,bbury=22.cm}}
\end{center}
\caption{} 
\label{tassoopal}
\end{figure}

\newpage

\begin{figure}[p]
\begin{center} 
\mbox{\epsfig{file=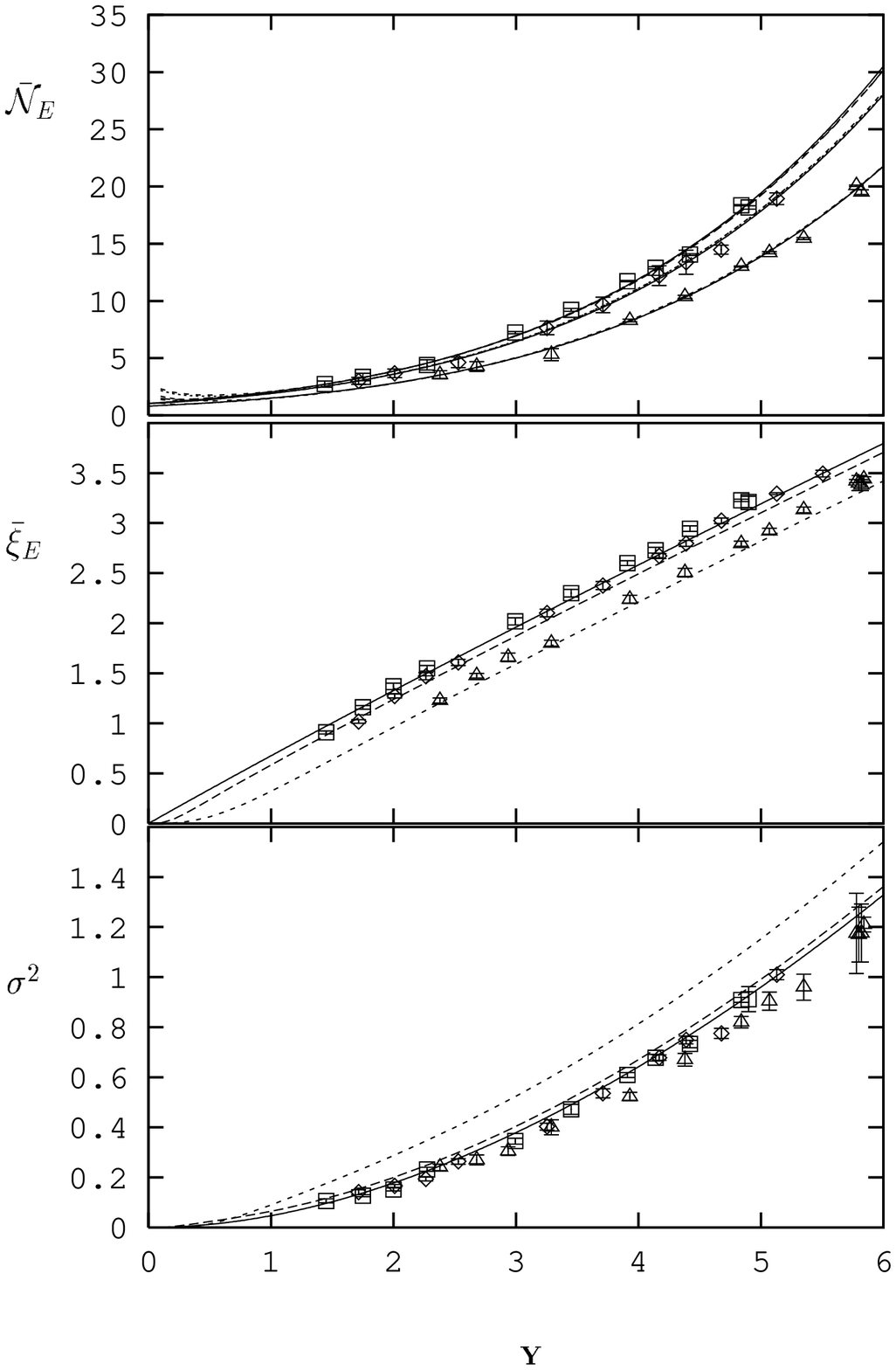,width=15cm,bbllx=4.cm,bblly=4.cm,bburx=19.cm,bbury=20.cm}}
\end{center}
\caption{} 
\label{test3r}
\end{figure}

\newpage 

\begin{figure}[p]
\begin{center} 
\mbox{\epsfig{file=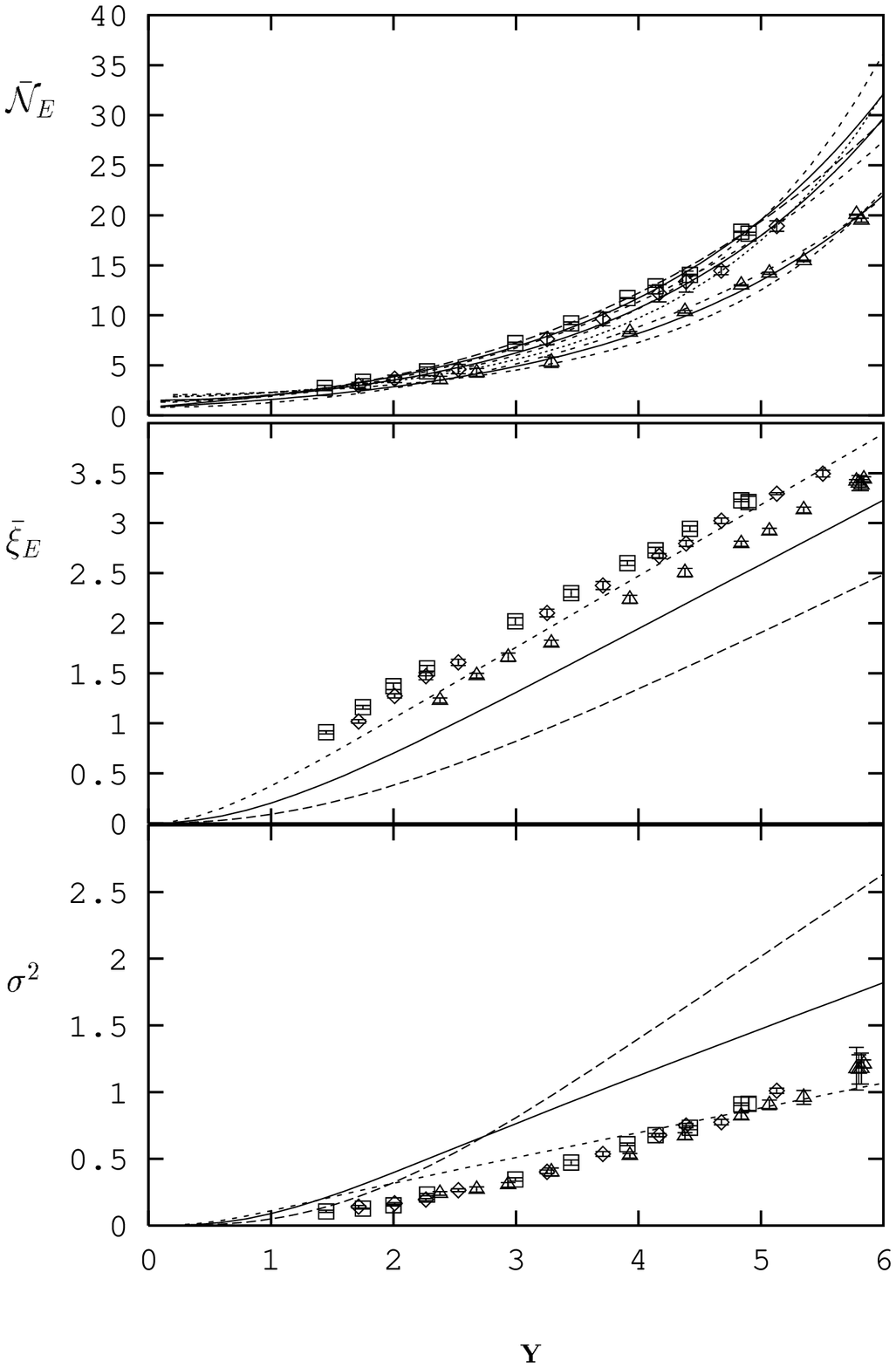,width=15cm,bbllx=4.cm,bblly=4.cm,bburx=19.cm,bbury=20.cm}}
\end{center}
\caption{} 
\label{test3f}
\end{figure}

\newpage 

\begin{figure}[p]
\begin{center} 
\mbox{\epsfig{file=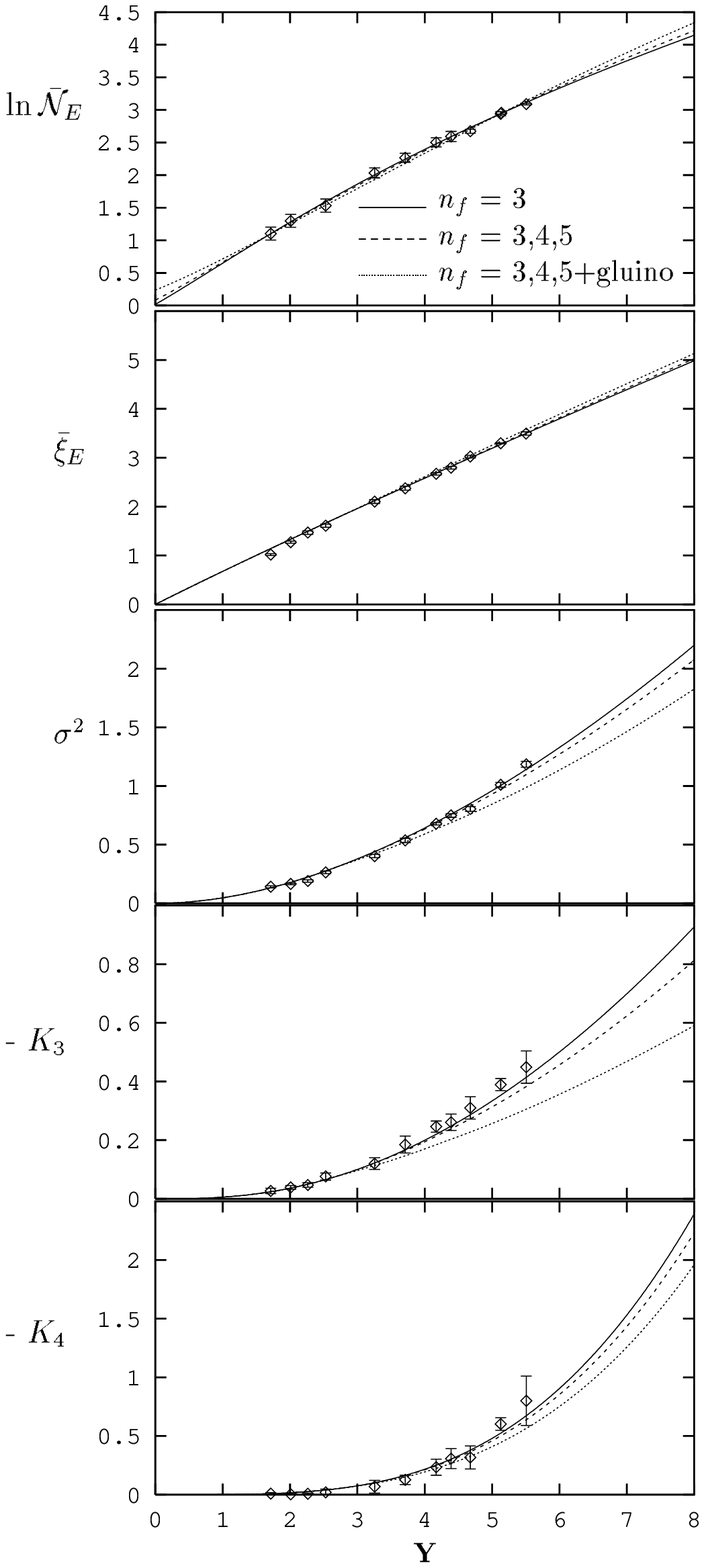,width=15cm,bbllx=4.5cm,bblly=2.cm,bburx=19.cm,bbury=23.cm}}
\end{center}
\caption{} 
\label{momentsflavour}
\end{figure}

\newpage 

\begin{figure}[p]
          \begin{center}
\mbox{\epsfig{file=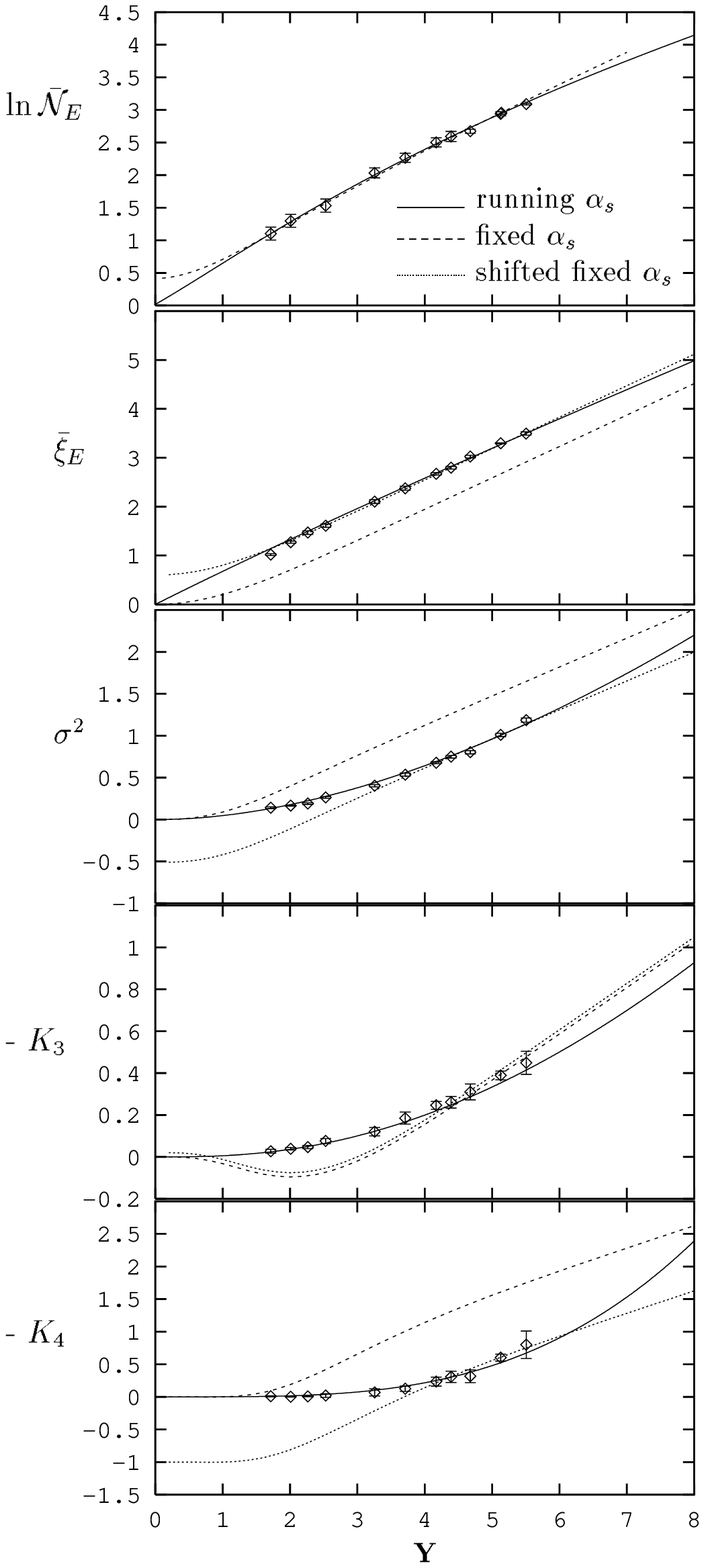,width=15.cm,bbllx=4.5cm,bblly=2.cm,bburx=19.cm,bbury=23.cm}}    
\end{center}
\caption{} 
\label{figmoments}
\end{figure}

\newpage

\begin{figure}[p]
\begin{center} 
\mbox{\epsfig{file=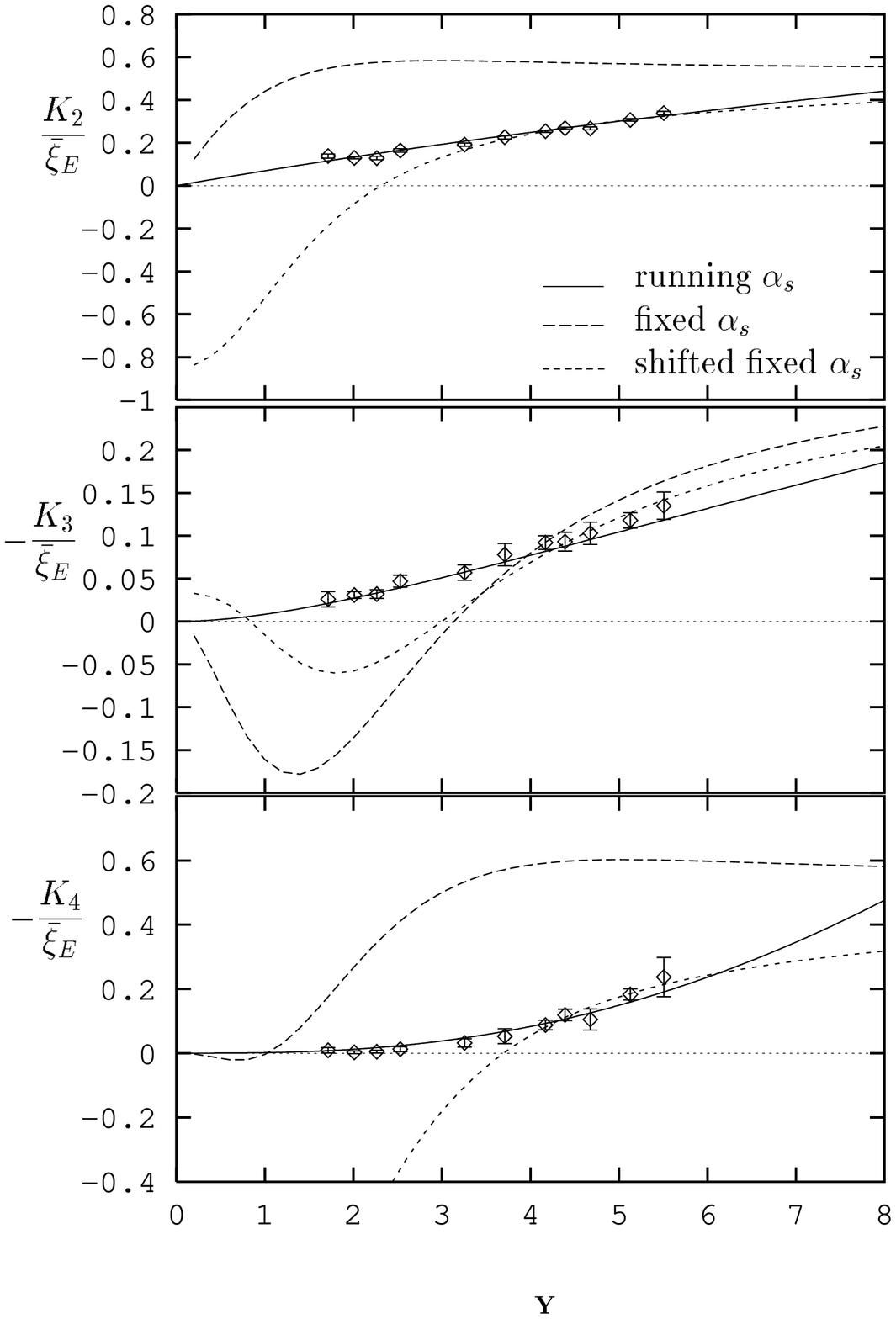,width=15cm,bbllx=4.cm,bblly=4.cm,bburx=19.cm,bbury=20.cm}}
\end{center}
\caption{} 
\label{reduced}
\end{figure}

\newpage

\begin{figure}[p]
          \begin{center}
\mbox{\epsfig{file=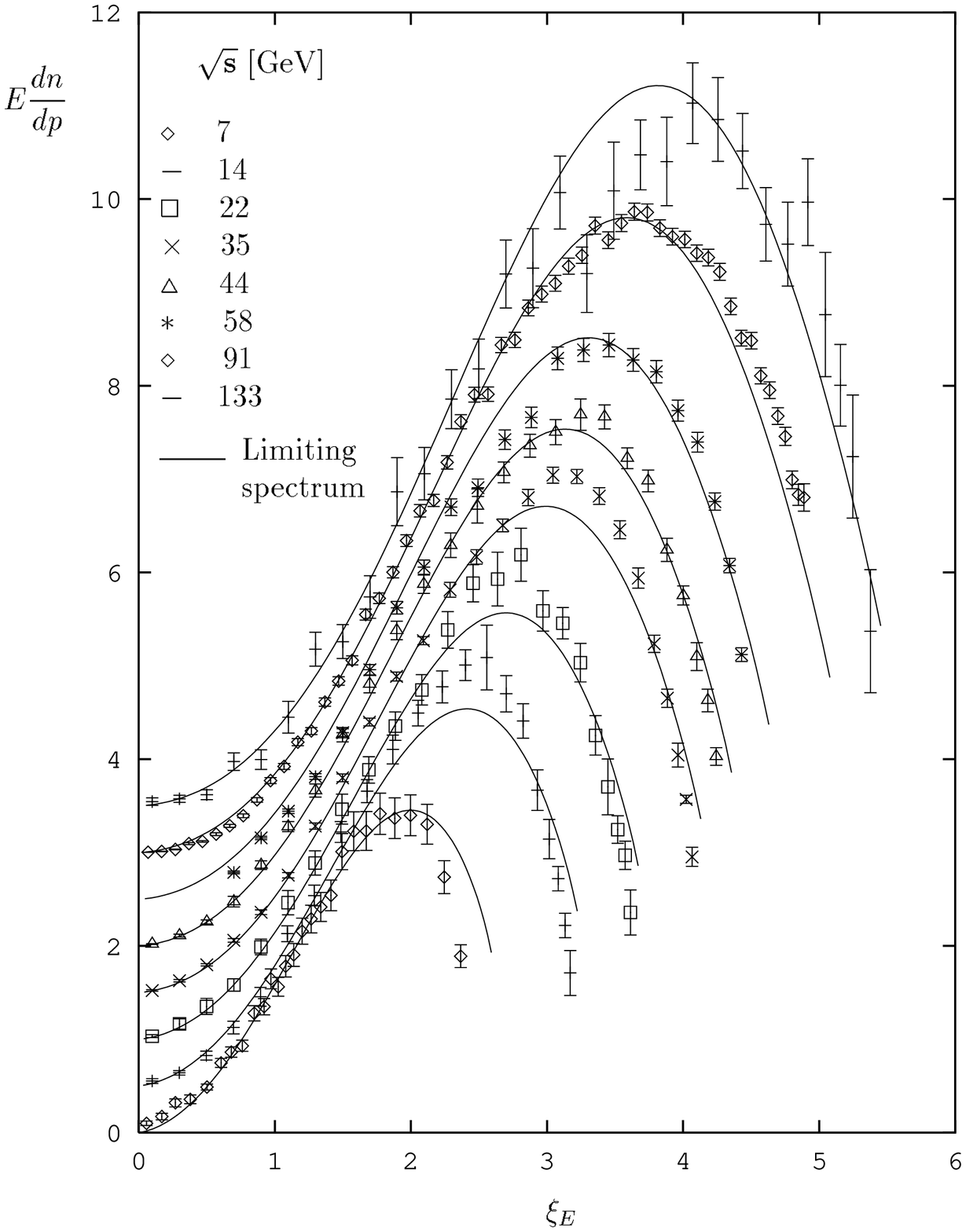,width=15.cm,bbllx=4.cm,bblly=5.cm,bburx=19.cm,bbury=25.cm}}    
\end{center}
\caption{} 
\label{shift}
\end{figure}

\newpage 

\begin{figure}[p]
          \begin{center}
\mbox{\epsfig{file=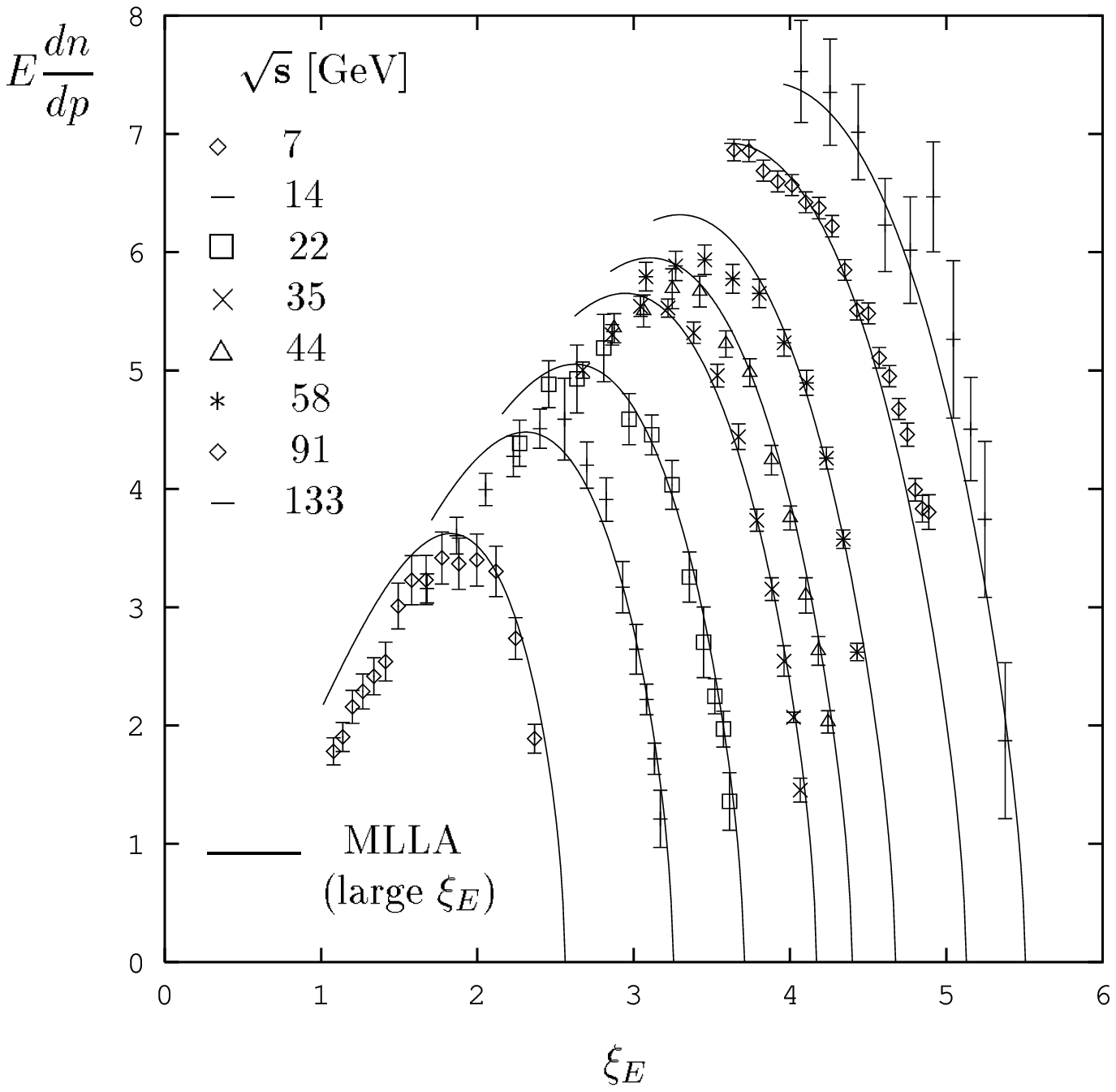,width=15.cm,bbllx=4.cm,bblly=5.cm,bburx=19.cm,bbury=25.cm}}    
\end{center}
\caption{} 
\label{itera}
\end{figure}

\newpage 

\begin{figure}[p]
\begin{center} 
\mbox{\epsfig{file=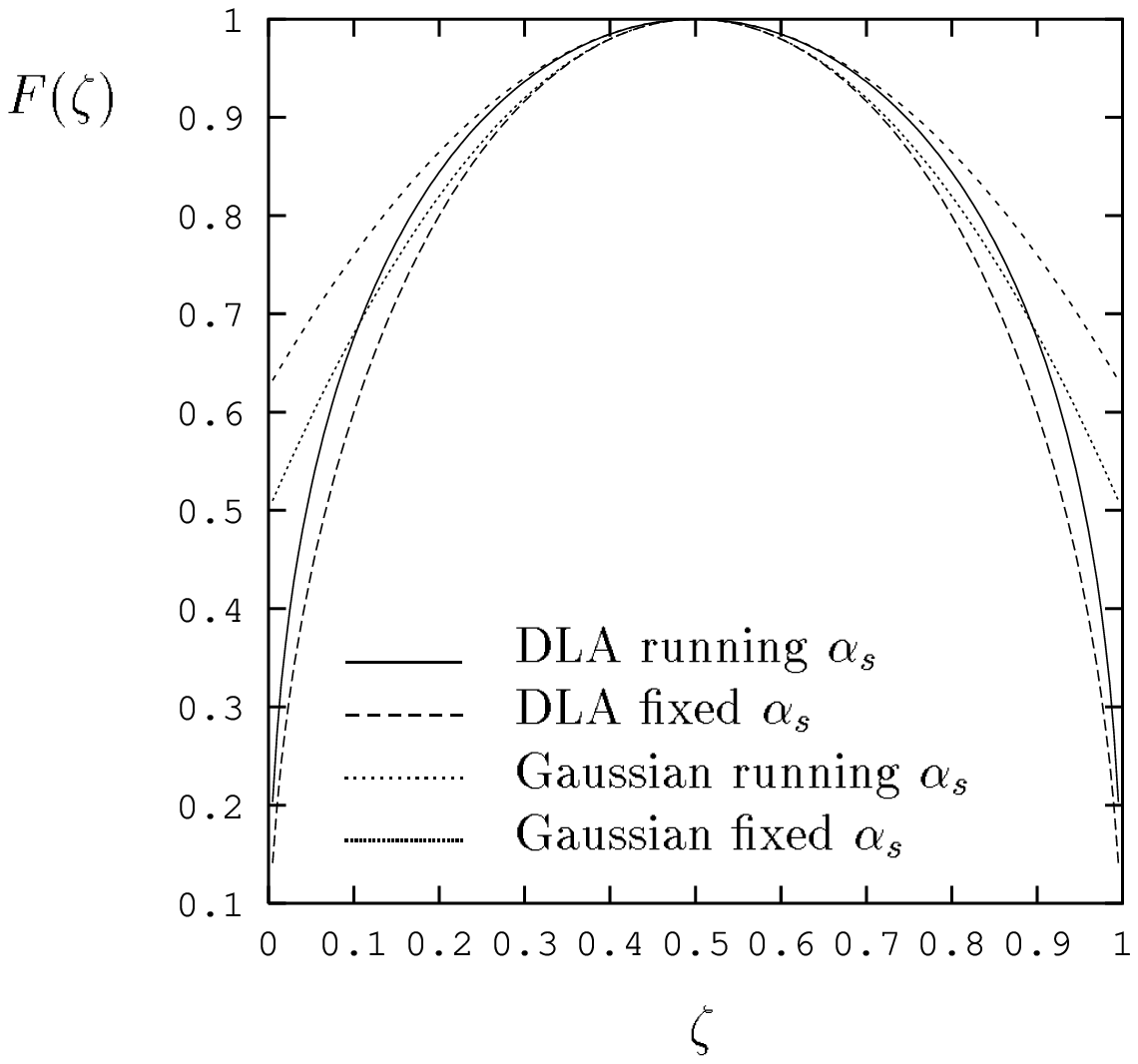,width=15cm,bbllx=4.cm,bblly=6.cm,bburx=19.cm,bbury=18.cm}}
\end{center}
\caption{} 
\label{zetatheory}
\end{figure}

\newpage 

\begin{figure}[p]
\begin{center} 
\mbox{\epsfig{file=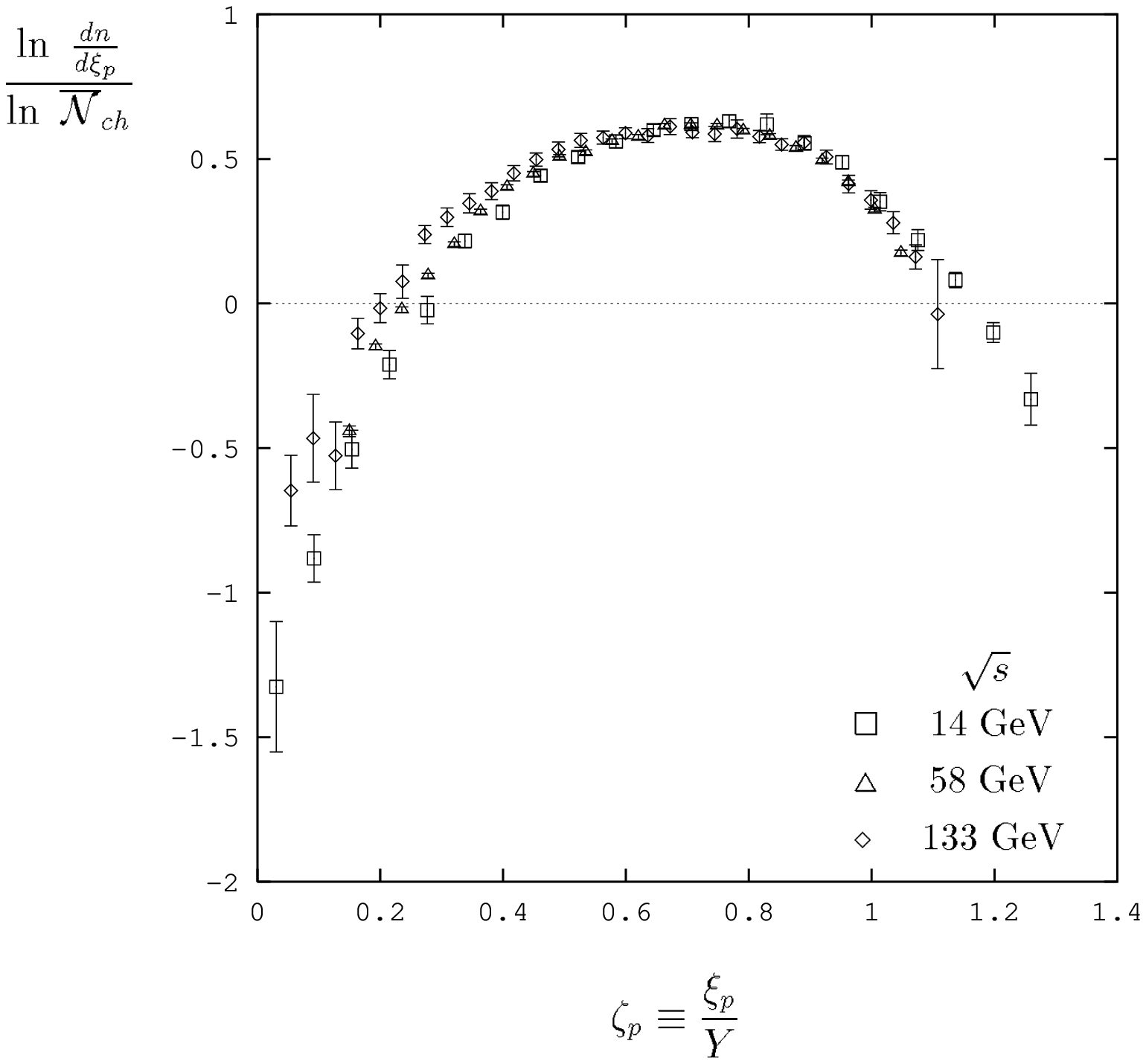,width=15cm,bbllx=4.cm,bblly=6.cm,bburx=19.cm,bbury=18.cm}}
\end{center}
\caption{} 
\label{zeta}
\end{figure}

\newpage 

\begin{figure}[p]
\begin{center} 
\mbox{\epsfig{file=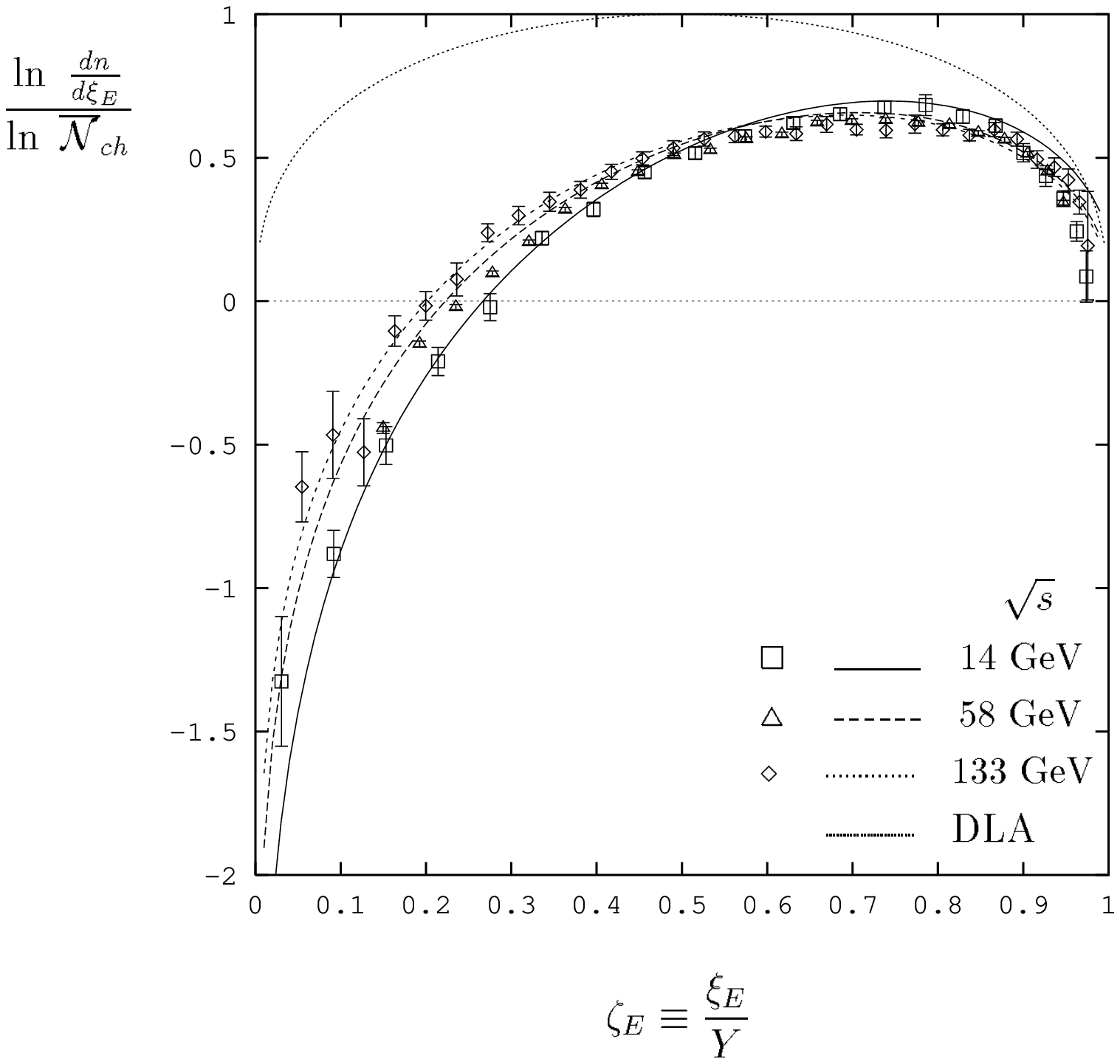,width=15cm,bbllx=4.cm,bblly=6.cm,bburx=19.cm,bbury=18.cm}}
\end{center}
\caption{} 
\label{zetae}
\end{figure}

\newpage 

\begin{figure}[p]
          \begin{center}
\mbox{\epsfig{file=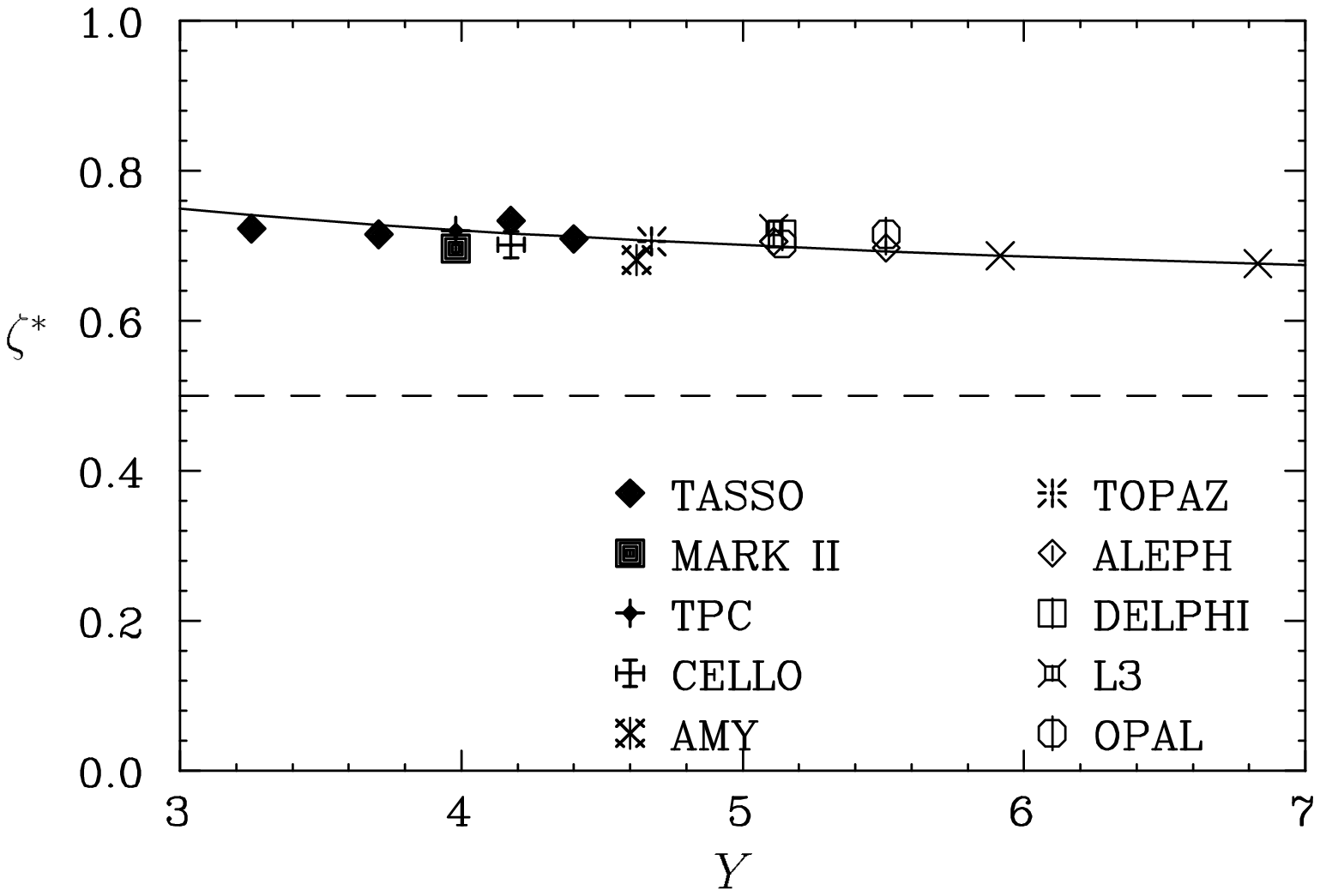,bbllx=2.8cm,bblly=14.cm,%
bburx=20.cm,bbury=26.cm,height=12.cm}}
          \end{center}
\caption{} 
\label{zetamax}
\end{figure}


\newpage 

\begin{thebibliography}{99}

\bibitem{BAS} A. Bassetto, M. Ciafaloni, G. Marchesini, 
Phys. Rep. {\bf 100} (1983) 202.

\bibitem{DKMTbook}
Yu. L. Dokshitzer, V.A. Khoze, A.H. Mueller, S.I. Troyan,
     Rev. Mod. Phys.  {\bf 60} (1988) 373; {\it ``Basics of
    Perturbative QCD''}, Editions Fronti\`eres, Gif-sur-Yvette
    CEDEX-France (1991).

\bibitem{LPHD} 
Ya. I. Azimov, Yu. L. Dokshitzer, V. A. Khoze, S. I. Troyan,
    Z. Phys.  {\bf C27} (1985) 65;  Z. Phys.  {\bf C31} (1986) 213. 

\bibitem{ko} 
V. A. Khoze and W. Ochs, ``Perturbative QCD 
approach to multiparticle production'',
preprint Durham DTP/96/36, MPI-PhT-96/29 (1996), to appear in Int. J. Mod Phys.
A.



\bibitem{DFK} 
Yu. L. Dokshitzer, V. S. Fadin, V. A. Khoze, Phys. Lett. {\bf B115} (1982) 242;
 Z. Phys.  {\bf C15} (1982) 325. 

\bibitem{MUEL} 
A. Bassetto, M. Ciafaloni, G. Marchesini and A. H. Mueller, 
Nucl. Phys. {\bf B207} (1982) 189. 

\bibitem{FW}
C. P. Fong and B. R. Webber, Nucl. Phys.  {\bf B355} (1991) 54.


\bibitem{ef} 
B. I. Ermolaev and V. S. Fadin, JETP Lett. {\bf 33} (1981) 285. 

\bibitem{ahm1} 
A.  H. Mueller, Phys. Lett. {\bf B104} (1981) 161. 

\bibitem{tasso}
TASSO Coll., W. Braunschweig et al., Z. Phys.  {\bf C47} (1990) 187. 

\bibitem{opal}
OPAL Coll., M. Z. Akrawy et al., Phys. Lett.  {\bf B247} (1990) 617. 

\bibitem{aleph15}
ALEPH Coll., D. Buskulic et al., Z. Phys.  {\bf C73} (1997) 409. 

\bibitem{delphi15}
DELPHI Coll., P. Abreu et al., Phys. Lett.  {\bf B372} (1996) 172.


\bibitem{opal15}
OPAL Coll., G. Alexander et al., Z. Phys.  {\bf C72} (1996) 191.


\bibitem{klo1} 
V. A. Khoze, S. Lupia and W. Ochs, Phys. Lett. {\bf B386} (1996) 451.

\bibitem{Korytov} 
CDF Coll., A. Korytov, in Proc. of QCD '96, 
(Montpellier, France, 1996), Eds. S. Narison,  
Nucl. Phys. B (Proc. Suppl.) {\bf 54A} (1997) 67.

\bibitem{DKTInt}
Yu. L. Dokshitzer, V. A. Khoze and S. I. Troyan, Int. J. Mod. Phys.  {\bf A7} (1992)
1875.

\bibitem{lo} 
S. Lupia and W. Ochs, Phys. Lett.  {\bf B365} (1996) 339.

\bibitem{conf} 
S. Lupia, in Proc. of QCD '96, 
(Montpellier, France, 1996), Eds. S. Narison,  
Nucl. Phys. B (Proc. Suppl.) {\bf 54A} (1997) 55; \\ 
W. Ochs, in Proc. XXXVI Cracow School of Theor. Phys., 
(Zakopane, Poland, 1996), Acta Phys. Pol. {\bf B27} (1996) 3505. 

\bibitem{dfk}   
 Yu.\ L.\ Dokshitzer, V.\ S.\ Fadin and V.\ A.\
Khoze, Z. Phys.  {\bf C18} (1983) 37. 

\bibitem{do} 
Yu. L. Dokshitzer and M. Olsson, Nucl. Phys.  {\bf B396} (1993) 137. 

\bibitem{dfk1}  
 Yu.\ L.\ Dokshitzer, V.\ S.\ Fadin and V.\ A.\
Khoze, Phys.\ Lett. {\bf B115} (1982) 242.

\bibitem{klo2} 
V. A. Khoze, S. Lupia and W. Ochs,  Phys. Lett. {\bf B394} (1997) 179. 




\bibitem{slac}
MARK I Coll., J. L. Siegrist et al., Phys. Rev.  {\bf D26} (1982) 969.  


\bibitem{slac2}
MARK II Coll., J. F. Patrick et al., Phys. Rev. Lett.  {\bf 49} (1982) 1232.  


\bibitem{topaz}
TOPAZ Coll., R. Itoh et al., Phys. Lett. {\bf B345} (1995) 335.

\bibitem{aleph}
ALEPH Coll., D. Buskulic et al., Z. Phys.  {\bf C55} (1992) 209. 

\bibitem{delphi}
DELPHI Coll., P. Abreu et al., Phys. Lett.  {\bf B347} (1995) 447.


\bibitem{l3}
L3 Coll., B. Adeva et al., Phys. Lett.  {\bf B259} (1991) 199. 


\bibitem{Marciano} 
W. Marciano, Phys. Rev. {\bf D29} (1983) 580; \\ 
Yu. L. Dokshitzer, D. V. Shirkov, Z. Phys. {\bf C67} (1995) 449. 

\bibitem{altarelli}
G. Altarelli, Phys. Rep. {\bf 81} (1982) 1. 


\bibitem{gluino} 
G. Farrar, Phys. Rev.  {\bf D51} (1995) 3904. 

\bibitem{alephgluino} 
ALEPH Coll., R. Barate et al., CERN-PPE/97-002, January 97, subm. to Z. Phys.
C.

\bibitem{ellisold} 
I. Antoniadis, J. Ellis and D. V. Nanopoulos, Phys. Lett.  {\bf B262} (1991)
109. 


\bibitem{cuypers} 
F. Cuypers, Z. Phys.  {\bf C61} (1994) 607. 

\bibitem{botterweck} 
F. Botterweck, private communication.  

\bibitem{jetset} 
T. Sj\"ostrand, Computer Physics Commun. {\bf 82} (1994) 74.


\bibitem{h1} 
H1 Coll., S.~Aid et al., Nucl.~Phys.~{\bf B445} (1995) 3.

\bibitem{zeus} 
ZEUS Coll.,  M.~Derrick et al., Z.~Phys.~{\bf C67} (1995) 93.


\bibitem{KNO}   
 Z.\ Koba, H. B.\ Nielsen and P.\ Olesen, Nucl. Phys. {\bf B40} (1972) 317.

\bibitem{poly}   
 A. M.\ Polyakov,  Sov.\ Phys.\ JETP {\bf 32} (1971) 296; ibid. 
 {\bf 33} (1971) 850.

\bibitem{bcm1}   
 A.\ Bassetto, M.\ Ciafaloni, G.\ Marchesini,  Nucl.\
Phys. {\bf B163} (1980) 477.

\bibitem{ow}   
W.\ Ochs and J.\ Wosiek, Phys.\ Lett.  {\bf B304} (1992) 144;
Z.\ Phys.\ {\bf C68} (1995) 269.

\bibitem{bm}   
DELPHI Coll.,  B.\ Buschbeck, F.\ Mandl\ et al.,
Proc.\ XXIV Int.\ Symp.\
on Multiparticle Dynamics, Sept. 1994, Vietri sul Mare, Salerno, Italy,
Eds.\ A.\ Giovannini, S.Lupia and R. Ugoccioni, World Scientific,
Singapore (1995) 52.


\bibitem{tassomult} 
TASSO Coll., W. Braunschweig et al., Z. Phys.  {\bf C45} (1989) 193. 

\bibitem{topazmult} 
TOPAZ Coll., M. Yamauchi et al., Proc. XXIV Int. Conf. on High Energy Physics 
(Munich, 1988), Springer Verlag, Eds. R. Kotthaus and J. H. K\"uhn, p.~852. 

\bibitem{mw}   
 E.\ D.\ Malaza, B.\ R.\ Webber,  Phys.\ Lett. 
{\bf B149} (1984) 501; Nucl.\ Phys. {\bf B267} (1986) 70.

\bibitem{yld}   
 Yu.\ L.\ Dokshitzer,  Phys.\ Lett. {\bf B305} (1993)
295.

\bibitem{ugl}   
R.\ Ugoccioni, A.\ Giovannini and S.\ Lupia, XXIV Int.\
Symp.\ on Multiparticle Dynamics 1994, Vietri sul Mare, Italy, Sept.\ 1994,
Eds.\ A.\ Giovannini et al., World Scientific, Singapore, p.\ 384 (1995).





\end{thebibliography}
\end{document}